\documentclass[structabstract]{aa}  
\usepackage{graphicx}
\usepackage{txfonts}
\usepackage{natbib}
\usepackage{longtable}
\usepackage{longtable,lscape}
\usepackage{color}
\begin{document}

   \title{FORS2/VLT survey of Milky Way globular clusters}
   \subtitle{II. Fe and Mg abundances of 51 Milky Way globular clusters
     on a homogeneous scale \thanks{Based on
       observations collected at the European Southern
       Observatory/Paranal, Chile, under programmes 68.B-0482(A),
       69.D-0455(A), 71.D-0219(A), 077.D-0775(A), and 089.D-0493(B).}}

   \author{B. Dias\inst{1,2,3}
          \and B. Barbuy\inst{2}
         \and I. Saviane\inst{1}
          \and E. V. Held\inst{4}
          \and G. S. Da Costa\inst{5}
          \and S. Ortolani\inst{4,6}
          \and M. Gullieuszik\inst{4}
          \and S. V\'asquez\inst{7,8,9}
%          \fnmsep
          }
   \institute{European Southern Observatory, Alonso de Cordova 3107,
         Santiago, Chile \\ \email{bdias@eso.org}
         \and Universidade de S\~ao Paulo, Dept. de Astronomia, Rua do Mat\~ao 
     1226, S\~ao Paulo 05508-090, Brazil
         \and Department of Physics, Durham University, South Road,
         Durham DH1 3LE, UK
         \and INAF, Osservatorio Astronomico di Padova, Vicolo dell'Osservatorio 5,
35122 Padova, Italy
         \and Research School of Astronomy \& Astrophysics, Australian National University, Mount Stromlo Observatory,
    via Cotter Road, Weston Creek, ACT 2611, Australia 
         \and Universit\`a di Padova, Dipartimento di Astronomia, Vicolo
 dell'Osservatorio 2, 35122 Padova, Italy
         \and Instituto de Astrofísica, Pontificia Universidad Católica de Chile, Av. Vicu\~na Mackenna 4860, 782-0436 Macul, Santiago, Chile
         \and Millennium Institute of Astrophysics, Av. Vicu\~na Mackenna 4860, 782-0436 Macul, Santiago, Chile
         \and Museo Interactivo Mirador, Dirección de Educación, Av. Punta Arenas, 6711 La Granja, Santiago, Chile
             }

   \date{Received: ; accepted: }

% \abstract{}{}{}{}{} 
% 5 {} token are mandatory
 
  \abstract
  {Globular clusters trace the formation and evolution of the Milky Way and surrounding galaxies, and outline their chemical enrichment history. To accomplish these tasks it is important to have large samples of clusters with homogeneous data and analysis to derive kinematics, chemical abundances, ages and locations. }
  {We  obtain homogeneous metallicities and $\alpha$-element enhancement for 51 Galactic bulge, disc, and halo globular clusters that are among the most distant and/or highly reddened in the Galaxy's globular cluster system. We also provide membership selection based on stellar radial velocities and atmospheric parameters. The implications of our results are discussed.}
   {We observed R$\sim$2000 spectra in the wavelength interval 456-586 nm for over 800 red giant stars in 51 Galactic globular clusters. We applied full spectrum fitting with the code ETOILE together with libraries of observed and synthetic spectra. We compared the mean abundances of all clusters with previous work and with field stars. We used the relation between mean metallicity and horizontal branch morphology defined by all clusters to select outliers for discussion.} 
   {[Fe/H], [Mg/Fe], and [$\alpha$/Fe] were derived in a consistent way for almost one-third of all Galactic globular clusters. We find our metallicities are comparable to those derived from high-resolution data to within $\sigma$ = 0.08~dex over the interval  --2.5 $<$ [Fe/H] $<$ 0.0.  Furthermore, a comparison of previous metallicity scales with our values  yields $\sigma < 0.16$~dex. We also find that the distribution of [Mg/Fe] and [$\alpha$/Fe]  with [Fe/H] for the 51 clusters follows the general trend exhibited by field stars. It is the first time that the following clusters have been included in a large sample of homogeneous stellar spectroscopic observations and metallicity derivation: BH~176, Djorg~2, Pal~10, NGC~6426, Lynga~7, and Terzan~8. In particular,  only photometric metallicities were available previously for the first three clusters, and the available metallicity for NGC~6426 was based on integrated spectroscopy and photometry.  Two other clusters, HP~1 and NGC~6558, are confirmed as candidates for the oldest globular clusters in the Milky Way.}
   {Stellar spectroscopy in the visible at R $\sim$ 2000 for a large sample of globular clusters is a robust and efficient way to trace the chemical evolution of the host galaxy and to detect interesting objects for follow-up at higher resolution and with forthcoming giant telescopes. The technique used here can also be applied to globular cluster systems in nearby galaxies with current instruments and to distant galaxies with the advent of ELTs. }


   \maketitle
%
%________________________________________________________________

\section{Introduction}

One of the most important questions about our Universe is, How did
galaxies form and evolve? 
A useful approach is to observe stars of different ages
and stellar populations that have imprinted in their kinematics
and chemical abundances the signatures of their formation period.
 Globular clusters are fossils tracing formation processes
 of their host galaxies at early epochs ($\sim$ 10-13~Gyr ago)
   and of more recent processes involving mergers with satellite galaxies.

Understanding the system of Galactic globular clusters (GGC) is of prime importance
to build up a picture of the formation and early evolution of the
Milky Way. Cluster ages are used to place the GGCs in the chronology
of our Galaxy;  the evolution of their chemical abundances and
kinematics provides evidence for the dynamical and chemical
evolution of the protogalactic halo and bulge. 

With the advent of multifibre spectrographs used in 8m class telescopes,
high-resolution spectra of sufficient S/N can now be obtained for the
cluster giant stars out to (m-M)$_V \approx$  19, i.e. for $>$
80\% of the GGCs. Nevertheless, the largest homogeneous samples
of metallicities are still based on low-resolution spectroscopy (and
calibrated with high spectral resolution results). Even so, more than 50\% of
GGCs do not have any spectroscopic estimation of their [Fe/H]
(see \citealp{saviane+12msn} for a review of [Fe/H]
values available in the literature).

Homogeneous determinations of [Fe/H] and [$\alpha$/Fe] for a large set
of globular clusters are useful to analyse the chemical evolution of
the different components of the Milky Way (bulge, disc, inner and
outer halo), and to allow comparisons with field stars. 
The combination of these
abundances with distance to the Galactic centre and ages leads to
discussions about the origin of globular clusters and
constrains models of the Galaxy's formation and evolution. In such studies
the [Fe/H] values for the GGCs arise from different sources that use
different methods and spectral resolution, gathered together on a single
scale. This procedure is useful to draw an overall picture of the
metallicity distribution of our Galaxy, but has an inherent uncertainty because of
the inhomogeneity of the abundance determinations.

In this work we present metallicity
[Fe/H], [Mg/Fe], [$\alpha$/Fe], and radial velocities for 51 of
the 157 Galactic globular clusters in the Harris catalogue 
(Harris 1996,  updated in 2010) from mid-resolution stellar spectra
(R$\sim$2000). Our
survey targets are mostly distant and highly
reddened clusters, which are poorly studied in the literature. We also
observed some well-known clusters for validation purposes
The sample was observed
with the same set-up,  analysed in a homogeneous way, and 
validated by comparing the data with high-resolution results in a
  complementary way to the approach discussed in \citet[][hereafter
Paper I]{dias+15}. We discuss how these results
can help to understand the formation and evolution of the Milky
Way. Similar observations and analysis techniques can be used
to study extragalactic globular clusters, such as those in the Magellanic Clouds,
in dwarf galaxies and in more distant galaxies, particularly
with the emergence of
40m class telescopes, such as the E-ELT.
The method of analysis is described in detail in Paper I.

In Sect. \ref{sec:obs} the selection of targets and observations are
described. In Sect. \ref{sec:met} the method detailed in Paper I is
summarized. Results are presented and the [Fe/H] values are compared to previous
metallicity scales in Sect. \ref{sec:scale}. Chemical evolution of the
Milky Way is briefly discussed in Sect. \ref{chem-MW}. 
In Sect. \ref{HB-sect} the second
parameter problem for horizontal branch morphology is considered and used to
select candidates for the oldest globular clusters in the
Galaxy. Finally, a summary and conclusions are given in
Sect. \ref{sec:conc}. 

%
%________________________________________________________________

\section{Target selection and observations}
\label{sec:obs}

\begin{figure}[!htb]
\centering
\includegraphics[width=\columnwidth]{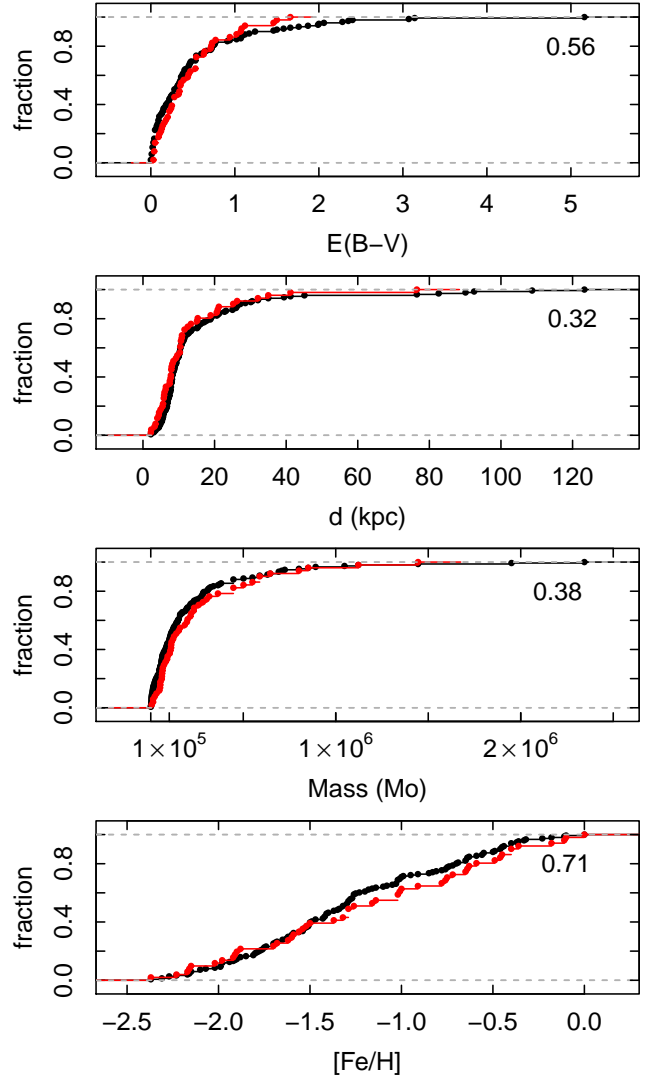}
\caption{Cumulative distributions of reddening, distance, stellar mass, and
  metallicity for Galactic globular clusters in the top to bottom panels, respectively.
  Black curves represent the total sample (157 clusters) from the catalogue
  of \citet[][2010 edition]{harris96},
  and red curves are the sample from this work (51 clusters). The
  numbers in the panels are the p-values obtained by applying the Kolmogorov-Smirnov
  test to the black and red curves.}
\label{sampl}
\end{figure}

Half of the targets were selected from the globular clusters catalogued
by \citet[][2010 edition]{harris96}\footnote{physwww.mcmaster.ca/$\sim$harris/mwgc.dat} 
that are more distant and/or highly reddened; many of them are poorly studied. 
The other half of  the sample consists of well-known brighter objects,
observed for comparison with high-resolution spectroscopic studies available in the literature. 
In Fig. \ref{sampl} we show the cumulative distribution of our
  sample of 51 clusters in comparison with the total sample from the catalogue of
\citet[][2010 edition]{harris96}, in terms of reddening, distance, stellar mass, and
  metallicity. Masses were
calculated by \cite{norris+14}. When not available we estimated masses from
the $M_{\rm V}$-$M_*$ relation (Eq. \ref{eq-norris}) that we fitted
from the \cite{norris+14} data and applied to $M_{\rm V}$ from
\citet[][2010 edition]{harris96}:

\begin{equation}
ln(M_*/M_{\odot}) = 4.33 -1.008 \cdot M_V
\label{eq-norris}
.\end{equation}

  \noindent The shapes of the distributions are very similar. For an
  objective comparison, we ran Kolmogorov-Smirnov tests and all
  p-values are greater than 5\%, meaning that the
  distributions are probably drawn from the same underlying population. 
  In other words, our sample is a 
  bias-free representation of the Milky Way globular
  cluster system.

\begin{figure*}[!htb]
\centering
\includegraphics[height=\textwidth,angle=-90]{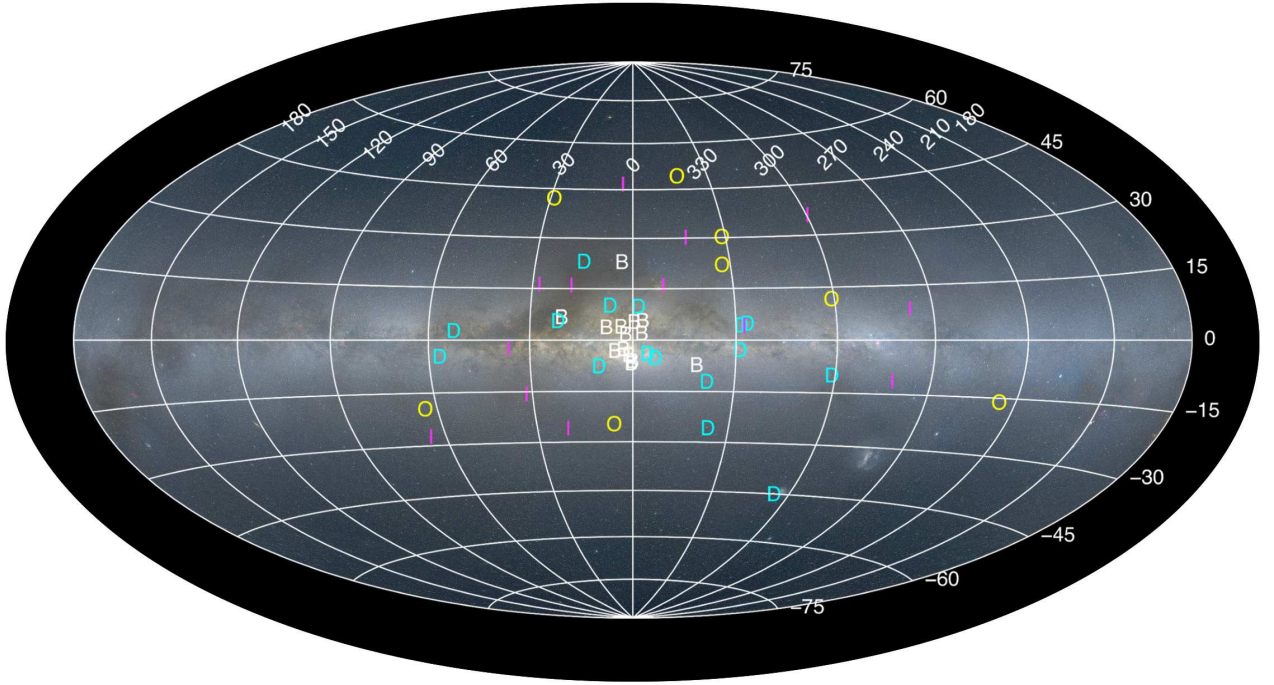}
\caption{Sky
  positions of the 51 clusters studied in this work over a Milky Way image \citep{mellinger09}
in terms of Galactic coordinates in an Aitoff projection. At the
position of each cluster, a letter indicates its Galactic component, namely
(B)ulge, (D)isc, (I)nner halo, and (O)uter halo, as given in Table
\ref{logtable}.}
\label{mwgc-aitoff}
\end{figure*}

The sample clusters
were subdivided into the four Galactic components (bulge, disc, inner
halo, and outer halo) following the criteria discussed by
\cite{carretta+10}, except for bulge clusters that were classified in
more detail by \cite{bica+15}, where a selection by
angular distances below 20$^{\circ}$ of the Galactic center,
galactocentric distances R$_{\rm GC} \leq 3.0$~kpc, and [Fe/H] $>$ -1.5
was found to best isolate bona fide bulge clusters.
According to Carretta et al.,
outer halo clusters have R$_{\rm GC} \geq 15.0$~kpc; the other objects
are classified as disc or inner halo depending on their kinematics
(dispersion or rotation dominated) and vertical distance with respect to
 the Galactic plane (see \citealp{carretta+10} for further
details). The  classification we adopted for each cluster is explicitly
shown in Table \ref{logtable} together with   the classifications of
Carretta et al. and Bica et al. We assigned the clusters classified as
non-bulge by Bica et al.  to the disc cluster category,
except for Pal~11, which is classified as inner halo by Carretta et al.
The sky positions of our sample of clusters, categorized by Galactic
  component, are displayed over an all-sky
image\footnote{https://sites.google.com/a/astro.ufrj.br/astronomer/home/allsky-projections-in-r}
in Figure \ref{mwgc-aitoff}.

\begin{figure}[!htb]
\centering
\includegraphics[width=\columnwidth]{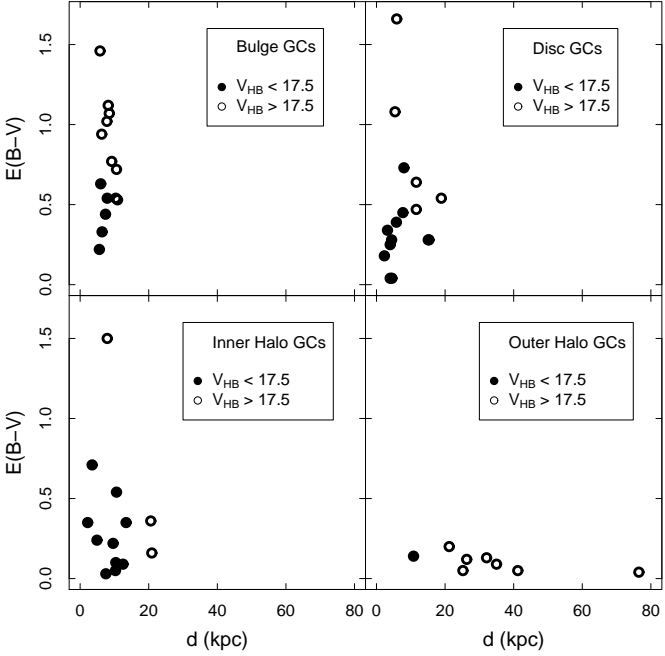}
\caption{Reddening versus distance of the 51 Milky Way globular clusters in our sample separated by
  Galactic component: bulge, disc, inner halo, and outer halo. Empty circles
  represent clusters with horizontal branch magnitude fainter than V =
  17.5, filled circles are clusters with V$_{\rm HB}$ brighter than 17.5.} 
\label{ebv-dist}
\end{figure}

Figure \ref{ebv-dist} shows reddening versus distance for our clusters
in the bulge, disc, inner, and outer halo subsamples.  The bulge clusters are located
at similar distances from the Sun ($\sim$~8~kpc) and are spread over
a wide range of reddening values between $\sim$~0.2 and 1.5 depending
on the direction. The closest disc clusters to the Sun have distances
of $\sim$~2.3~kpc and the farthest are $\sim$~19~kpc from
the Sun. Because of this distribution and the low latitudes of these
clusters, reddening values vary from $\sim$~0.04 to
$\sim$~1.7. The inner halo objects have similar
 intervals of E(B-V) and distance to  the
disc clusters. Outer halo clusters have low reddening, E(B-V)~$<$~0.2
and are far from the Sun (11~$<$~d(kpc)~$<$~77). 

To have homogeneous observations for all targets we chose the
  multi-object spectrograph, the FORS2 instrument on board the ESO Very Large Telescope (FORS2@VLT/ESO)\citep{appenzeller+98}. This instrument enables
  observations of the faintest and brightest stars in our sample
with a good compromise between signal-to-noise and exposure time. For
example, the faintest stars we observed have V $\approx$ 19, and one hour
of exposure with FORS2\footnote{Exposure time calculator,
  http://www.eso.org/observing/etc/} results in S/N $\sim$ 50 at this magnitude, which is
sufficient for our techniques. Higher resolution spectrographs such as
the FLAMES instrument on board the ESO Very Large Telescope (FLAMES@VLT/ESO)
 require prohibitive amounts of telescope time for stars of this 
faint magnitude.  Specifically, the
FLAMES user manual\footnote{FLAMES User
  Manual VLT-MAN-ESO-13700-2994, Issue 92, 06/03/2013, Table 1}
indicates that observations of stars with V=17.5 with one hour of
exposure will produce spectra with S/N $\sim$ 30 using GIRAFFE fibres and S/N
$\sim$ 10 using UVES fibres.  Stars fainter than that would require too
much telescope time to obtain useful spectra. Consequently, detailed
abundance studies based on optical spectra of stars in the more distant/reddened clusters 
are not feasible at the present time, and must await future Extremely Large Telescope (ELT) class facilities.

We selected red giant stars usually brighter than the
horizontal branch level (see Paper I) in each cluster;
therefore, the classification of globular clusters with V$_{\rm HB}
>$ 17.5 indicates which clusters could not be observed with
high-resolution optical spectroscopy with current facilities. 
In Figure \ref{ebv-dist} we note  that half of the sample clusters (25 of 51) across all Galactic components are not
bright enough for high-resolution observations. Thus, our survey
represents a significant improvement in our knowledge of the chemical
content of Milky Way globular clusters. 
We also note that 13 of the brightest clusters are in common
with observations that defined the metallicity scale of 
\cite{carretta+09}. In Sect. \ref{sec:scale} we compare in more detail
our results with previous metallicity scales.

Around 16 red giant stars were selected from photometry for each cluster from 
the pre-imaging observations for a total of 819 stars.
We obtained FORS2@VLT/ESO spectra for them, and
61 spectra (7\%) were not considered in the analysis owing to very low 
signal-to-noise ratio or data reduction problems. From the remaining 
758 useful spectra, 465 (61\%) are of confirmed member stars of the 51
clusters.
The spectra were observed in the visible region (grism 1400V, 456 -
586~nm) with resolution of $\Delta\lambda$=2.5\rm{\AA} and typical
S/N$\sim$30 - 100. The data were collected from 2001 to 2012 under
projects ID 68.B-0482(A, 2001), ID 69.D-0455(A, 2002), ID 71.D-0219(A,
2003), ID 077.D-0775(A, 2006), and ID 089.D-0493(B, 2012).
Table \ref{logtable} lists the selected clusters, their coordinates,
observing dates, and exposure times. Coordinates of the 758
analysed stars and their magnitudes are given in Table
\ref{starinfo}.
The spectra were reduced using FORS2 pipeline inside the {\it EsoRex}
software\footnote{https://www.eso.org/sci/software/cpl/esorex.html}
following the procedure described in Paper I.

%% Table \ref{logtable}. #1
%% Log of observations, population, number of member/non-member stars
\addtocounter{table}{1}

\begin{table*}[!htb]
\caption{Identifications, coordinates,
  instrumental magnitudes and colours, and heliocentric radial
  velocities for all the stars observed. Velocities from CaT were taken from \cite{saviane+12}
  and from Vasquez et al. (in prep.)}
\label{starinfo} 
\centering
\begin{tabular}{lllcccrrc}
\hline\hline
\noalign{\smallskip}
 {\rm Star ID} & {\rm RA (J2000)} & {\rm DEC (J2000)} & {V$_{\rm
     instr.}$} & {(V-I)$_{\rm instr.}$} & {(B-V)$_{\rm instr.}$} & {\rm v$_{\rm helio}$} & {\rm
   v$_{\rm helio-CaT}$} & members\\
 & (hh:mm:ss)  & (dd:mm:ss) & (mag) & (mag) & (mag) & {\rm (km/s)} & {\rm (km/s)} & \\
\noalign{\smallskip}
\hline
\noalign{\smallskip}
47Tuc\_502 &    00:25:35         & -72:02:23  &  11.40   &  ---  & 2.66  &  -72.61  &  ---  &      \\    
47Tuc\_509 &    00:25:18         & -72:05:03  &  12.09   &  ---  & 1.47  &  -35.21  &  ---  &  M     \\
47Tuc\_514 &    00:25:41         & -72:05:57  &  13.74   &  ---  & 1.04  &  -46.80  &  ---  &  M     \\
47Tuc\_517 &    00:25:41         & -72:06:25  &  13.44   &  ---  & 1.16  &  -54.61  &  ---  &  M     \\
47Tuc\_519 &    00:25:40         & -72:06:34  &  12.69   &  ---  & 1.28  &  -56.79  &  ---  &  M     \\
47Tuc\_525 &    00:25:28         & -72:01:38  &  12.84   &  ---  & 1.27  &  -44.54  &  ---  &  M     \\
47Tuc\_533 &    00:25:20         & -72:05:16  &  12.09   &  ---  & 1.40  &  -34.56  &  ---  &  M     \\
47Tuc\_534 &    00:25:15         & -72:01:54  &  14.05   &  ---  & 0.83  &  -50.90  &  ---  &  M     \\
47Tuc\_535 &    00:25:22         & -72:05:29  &  13.79   &  ---  & 1.06  &  -49.11  &  ---  &  M     \\
47Tuc\_539 &    00:25:28         & -72:03:21  &  14.68   &  ---  & 0.90  &  -32.26  &  ---  &  M     \\
47Tuc\_551 &    00:25:15         & -72:04:13  &  12.34   &  ---  & 1.40  &  -40.29  &  ---  &  M     \\
47Tuc\_553 &    00:25:35         & -72:04:26  &  14.15   &  ---  & 0.84  &  -46.89  &  ---  &  M     \\
47Tuc\_554 &    00:25:28         & -72:00:58  &  12.06   &  ---  & 1.57  &  -44.67  &  ---  &  M     \\
47Tuc\_559 &    00:25:37         & -72:00:40  &  13.98   &  ---  & 1.05  &  -68.65  &  ---  &  M     \\
47Tuc\_571 &    00:25:29         & -72:02:36  &  12.59   &  ---  & 1.26  &  -58.65  &  ---  &  M     \\
47Tuc\_581 &    00:25:38         & -72:01:24  &  11.82   &  ---  & 1.66  &  -56.46  &  ---  &   M     \\
\noalign{\smallskip}
...  &   ...  &  ...    &... &    ...   & ...    &    ...    &  ...   &   ...         \\
\noalign{\smallskip}
\hline
\end{tabular}
\tablefoot{Complete version of this table for all 758 stars is available online at VizieR.}
\end{table*}

%
%________________________________________________________________

\section{Method}
\label{sec:met}
The method for atmospheric parameter derivation was  described and
exhaustively discussed in Paper I, and can be summarized as follows.
Atmospheric parameters (T$_{eff}$, log($g$), [Fe/H], [Mg/Fe],
[$\alpha$/Fe]) were derived for each star by applying full spectrum
fitting through the code ETOILE (\citealp{katz+11} and
\citealp{katz01}). The code takes into account {a priori}
T$_{eff}$ and log($g$) intervals for red giant branch stars and carries out
a $\chi^2$ pixel-by-pixel fitting of a given target spectrum
to a set of template spectra. We chose two libraries of template
stellar spectra, one empirical (MILES, \citealp{sanchez-blazquez+06}) and one
synthetic \citep[][hereafter referred to as
Coelho]{coelho+05}. 

The library spectra are sorted by
similarity ($S$, proportional to $\chi^2$, see Paper I) to the target
spectrum and the parameters are calculated by taking 
the average of the parameters of the top $N$ template spectra. For the
Coelho library we adopted $N=10$, and for the MILES library
$N$ is defined such that $S(N)/S(1) \lesssim$~1.1. 
For each star, [Mg/Fe] is given by the 
MILES templates only; [$\alpha$/Fe] is given by the Coelho templates only; and
T$_{eff}$, log($g$), and [Fe/H] are the averages of the MILES and Coelho results.
Uncertainties of T$_{eff}$, log($g$), [Fe/H], [Mg/Fe], and [$\alpha$/Fe]
for each star are the standard deviation of the average of the top $N$
templates. It is difficult to estimate the correlation between
  the parameters because of the nature of the adopted analysis technique (see details in Paper I).
The uncertainties of the average of the MILES and Coelho
results for each star are calculated through conventional propagation,
as are  the uncertainties for the average [Fe/H] for the member
stars of each globular cluster.

We note that before running the comparison of a given target
spectrum with the reference spectra, two important steps are needed:
 convolving all the library spectra to the same resolution of the target spectrum,
and  correcting them for radial velocities, also measured  with the same
ETOILE code using a cross-correlation method with one template spectrum.
For detailed discussion of this method
and validation with well-known stars and high-resolution analysis, we
refer to Paper I.

Membership selection of stars for each cluster was done in two steps:
first,  by radial velocities and metallicities; second, by
proximity of temperature and surface gravity to reference
isochrones, which is independent of reddening. In this way we use all the derived atmospheric parameters as
input in the selection of member stars. Examples and a detailed
description are given in Paper I.

%
%________________________________________________________________

\section{Results and comparison with previous metallicity scales}
\label{sec:scale}

Atmospheric parameters for all 758 studied stars are presented in
Table \ref{finalparam} following the IDs from Table
\ref{starinfo}. We list T$_{\rm eff}$, log($g$), and [Fe/H] from both the
MILES and Coelho libraries, and the averages of these values are adopted as
our final parameters (see Paper I for a detailed justification of this procedure). Table
\ref{finalparam} also lists [Mg/Fe] from MILES and
[$\alpha$/Fe] from the Coelho library. The average of ${\rm [Fe/H]}$,
${\rm [Mg/Fe]}$, ${\rm [\alpha/Fe]}$, and $v_{\rm helio}$ for the 51
clusters based on their selected member stars are presented in Table
\ref{finalparamavgcalib}. Metallicities from the MILES and Coelho
libraries are given;    the average of these results is    our final abundance for the clusters.
We note that while there are some clusters that are known to possess sizeable
spreads in individual [Mg/Fe] values, as a result of the light element
chemical anomalies usually referred to as the O-Na anti-correlation,
in most cases the spread in [Mg/Fe] is small
(e.g. \citealp{carretta+09uves}; Fig. 6). Therefore, our approach of averaging
all the determinations for a given cluster should not substantially bias the
mean value. 

The previous largest abundance collection for Galactic globular
  clusters was done by \cite{pritzl+05} for 45 objects, containing [Fe/H], [Mg/Fe], [Si/Fe], [Ca/Fe], [Ti/Fe], [Y/Fe], [Ba/Fe], [La/Fe], and [Eu/Fe]. More recently \cite{roediger+14} has compiled chemical abundances from the literature for 41 globular clusters, including [Fe/H], [Mg/Fe], [C/Fe], [N/Fe], [Ca/Fe], [O/Fe], [Na/Fe], [Si/Fe], [Cr/Fe], and [Ti/Fe]. The caveats of these compilations are that they are based on heterogeneous data available in the literature, and the objects are mostly halo clusters for the Pritzl et al. sample.
Our results represent the first time that [Fe/H], [Mg/Fe], and [$\alpha$/Fe], derived
in a consistent way, are given for such a large sample of globular
clusters (51 objects); this number is almost one-third of the total number of 
catalogued clusters (157 as compiled by \citealp[][2010 edition]{harris96}), and includes all Milky Way components.

%% Table \ref{finalparam}. #3
%% Log of observations, population, number of member/non-member stars
\addtocounter{table}{1}

%% Table \ref{finalparamavgcalib}. #4
%% Log of observations, population, number of member/non-member stars
\addtocounter{table}{1}

In the following sections we compare our metallicity determinations
with five other works that report homogenous metallicities for at
least 16 Galactic globular clusters. We begin with the high-resolution 
study of \cite[][hereafter C09]{carretta+09}  described in Table \ref{tab:car09lit}.

%------------------------------------------------

\subsection{Carretta et al. (2009a) scale}
\label{feshcarscale}

Carretta et al. (2009a) reported a new metallicity scale for Milky Way
globular clusters based on their observations of 19 clusters with UVES
\citep{carretta+09uves} and GIRAFFE \citep{carretta+09gir} at
VLT/ESO. This scale superseded their previous scale
\citep{carretta+97}. In our survey there are 13 objects in common with
their sample covering the metallicity range -2.3 $<$ [Fe/H] $<$ -0.4, as
shown in Table \ref{tab:fesh-car09}. Carretta et al. added two metal-rich clusters, 
NGC~6553 and NGC~6528, with previous
high-resolution spectroscopy to increase
the metallicity range up to solar abundance;
specifically, they adopted
  the abundances from \cite{carretta+01} who showed that these
  clusters have similar metallicities, and that  NGC~6528 is slightly more metal-rich. They derived 
[Fe/H] = +0.07$\pm$0.10 for NGC~6528 and 
then offset the value [Fe/H] = -0.16$\pm$0.08 for NGC~6553 from \citet[][C99]{cohen+99} 
  to  [Fe/H] = --0.06$\pm$0.15, a value closer to the one they found for
  NGC~6528. However, the metallicity derived by C99 agrees
well with more recent work. For example, \citet[][M03]{melendez+03} and
\citet[][AB06]{alves-brito+06} derived [Fe/H] = --0.2$\pm$0.1 and [Fe/H] =
-0.20$\pm$0.02 for this cluster. Therefore, the original
  value of C99 for NGC~6553  should be retained. We adopt here the 
  weighted mean metallicity of C99, M03, and AB06 for NGC~6553. In the case of NGC~6528
  a more recent work derived [Fe/H] = --0.1 $\pm$ 0.2  \citep[][Z04]{zoccali+04}, and we took the
  weighted mean metallicity of the values from Z04 and C01 as our reference for NGC~6528.
All values are compiled in Table \ref{tab:fesh-car09}.

\begin{table}[!htb]
\caption{Average [Fe/H] from this work compared with the 13 globular clusters in common 
with C09. For the two metal-rich clusters we adopted the mean metallicities
from \citet[][C01]{carretta+01} and \citet[][Z04]{zoccali+04} for NGC~6528, and
from \citet[][C99]{cohen+99}, \citet[][M03]{melendez+03}, and \citet[][AB06]{alves-brito+06}
for NGC~6553.}
\label{tab:fesh-car09}
\centering
\begin{tabular}{l@{ }l@{ }c@{ }c@{ }c}
\hline\hline
\noalign{\smallskip}
Cluster  & Other    & [Fe/H] & [Fe/H] & ref.\\
 & names    & (average) & (lit.) & \\
\noalign{\smallskip}
\hline
\noalign{\smallskip}
NGC 104    &    47 Tuc     &     -0.71$\pm$0.04      &       -0.77$\pm$0.05          & C09      \\
NGC 2808   &               &     -1.06$\pm$0.05      &       -1.15$\pm$0.07          & C09      \\
NGC 3201   &               &     -1.51$\pm$0.03      &       -1.51$\pm$0.06          & C09      \\
NGC 4590   &    M 68       &     -2.20$\pm$0.05      &       -2.26$\pm$0.05          & C09      \\
NGC 5904   &    M 5        &     -1.25$\pm$0.05      &       -1.34$\pm$0.05          & C09      \\
NGC 6121   &    M 4        &     -1.01$\pm$0.05      &       -1.17$\pm$0.05          & C09      \\
NGC 6171   &    M 107      &     -0.95$\pm$0.09      &       -1.03$\pm$0.04          & C09      \\
NGC 6254   &    M 10       &     -1.56$\pm$0.04      &       -1.57$\pm$0.06          & C09      \\
NGC 6397   &               &     -2.07$\pm$0.03      &       -1.99$\pm$0.04          & C09      \\
NGC 6441   &               &     -0.41$\pm$0.07      &       -0.43$\pm$0.06          & C09      \\
NGC 6752   &               &     -1.57$\pm$0.07      &       -1.55$\pm$0.05          & C09      \\
NGC 6838   &    M 71       &     -0.63$\pm$0.06      &       -0.83$\pm$0.06          & C09      \\
NGC 7078   &    M 15       &     -2.23$\pm$0.02      &       -2.32$\pm$0.06          & C09      \\
\noalign{\smallskip}
NGC 6528   &       &               -0.13$\pm$0.07      &       -0.02$\pm$0.09         & $<$C01,Z04$>$      \\
NGC 6553   &                      &     -0.13$\pm$0.01      &       -0.19$\pm$0.02          & $<$C99,M03,AB06$>$  \\
\noalign{\smallskip}
\hline
\end{tabular}
\end{table}
\normalsize

\begin{figure}[!htb]
\centering
\includegraphics[width=\columnwidth]{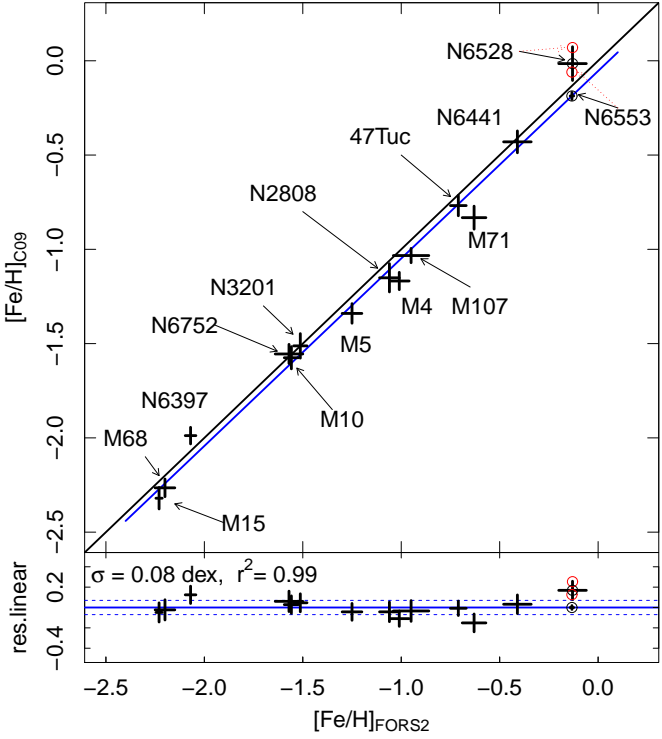}
\caption{Comparison of [Fe/H] from this work with those from
  C09 for the 13 clusters in common plus NGC~6553 from C01 and Z04, and
  NGC~6528 from C99, M03, and AB06. 
  The black line is the one-to-one relation and the blue line is the linear fit
to the 15 black points  (Eq. \ref{eq:fesh-car09}).
Residuals are presented in the bottom panels. The blue dashed lines represent $\pm 1\sigma$. 
Metallicities adopted by
 \cite{carretta+01} are shown in red for reference, but they are not considered
 in the fit. Values are listed in Table \ref{tab:fesh-car09}.}
\label{fig:fesh-car09}
\end{figure}

Our [Fe/H] results are compared with the 13 clusters from C09 plus the
two metal-rich clusters (see Table \ref{tab:fesh-car09}) averaged from other sources in Figure
\ref{fig:fesh-car09} where the cluster names are indicated. The
metal-rich clusters are indicated by circles and they are
included in the linear fit of Eq. \ref{eq:fesh-car09}, represented by
the blue line in the plot, and valid in the metallicity range -2.4 $<$
[Fe/H] $<$ 0.0: 

\begin{equation}
{\rm [Fe/H]}_{\rm C09} = -0.05(\pm0.04) + 0.99(\pm0.03){\rm
  [Fe/H]}_{\rm FORS2}
\label{eq:fesh-car09}
.\end{equation}

\noindent Metallicities of the two metal-rich clusters adopted by
\cite{carretta+01} are overplotted as red circles in Fig. \ref{fig:fesh-car09} for
reference, but they are not included in the fit. 

From Eq. \ref{eq:fesh-car09} we can conclude that our metallicity
results are in excellent agreement
with those from high-resolution spectroscopy
because the slope of the fit
 is compatible with 1.0 and the 
offset is near zero. The correlation coefficient,
$r^2$ = 0.99, is
close to unity and there is no indication of any correlation
between the residuals and metallicity, which justifies the use of a linear relation.
The standard deviation, $\sigma$ = 0.08~dex, can be explained by the
uncertainties in the individual cluster metallicities (see Table
\ref{tab:fesh-car09}, where our abundances are an excerpt of Table
\ref{finalparamavgcalib}). Moreover, the residuals plot shows explicitly that
the [Fe/H] values adopted by \cite{carretta+01} for NGC~6528
and NGC~6553 are shifted upwards from the relation by at least 1$\sigma$
 with respect to our adopted values. The consistency of our
results with C09  (complemented by metal-rich clusters from
other works based on high-resolution spectroscopy) scale in the
entire range -2.4 $<$ [Fe/H] $<$ 0.0 and
supports the robustness of the metallicities derived from full spectrum
fitting of low- or medium-resolution spectroscopy. 

We also note  that
  C09 used their adopted metallicities for NGC~6553 and NGC~6528
  in their recalibration of other metallicity scales. Since we have adopted
  lower metallicities for these clusters, values that agree well
  with other high-resolution spectroscopic work, the calibration of other metallicity
  scales -- particularly for the metal-rich tail -- needs to be reconsidered.

\subsection{Zinn \& West scale}
\begin{figure}[!htb]
\centering
\includegraphics[width=\columnwidth]{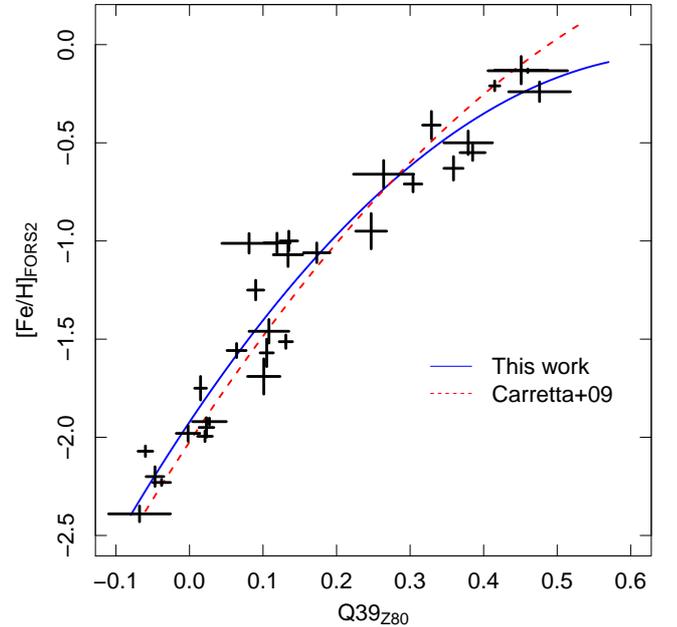}
\caption{$Q_{39}$ index from \cite{zinn80} against [Fe/H]
  from this work. The blue solid line is the quadratic function fitted to
  the data (Eq. \ref{eq:q39-d14}) and the red dashed line is the quadratic
  function fitted by C09 to calibrate $Q_{39}$ to their scale. Fitted
  curves are shown only in their respective valid ranges.}
\label{fig:q39-d14}
\end{figure}

\cite{zinn+84} published a metallicity scale 30 years ago that is
still a reference, although it is based on the integrated-light index
$Q_{39}$ \citep{zinn80}. We compare their $Q_{39}$ index with our final [Fe/H] values for
the 31 clusters in common in Figure \ref{fig:q39-d14}.  The relation is  described well by 
the second-order polynomial of Eq. \ref{eq:q39-d14}: 

\begin{equation}
\begin{array}{l}
{\rm [Fe/H]}_{\rm FORS2} = -1.92(\pm 0.05) + 5.6(\pm 0.6)\cdot Q_{39} -\\
-4.2(\pm 1.4)\cdot Q_{39}^2
\end{array}
\label{eq:q39-d14}
\end{equation}

\noindent The fitting quality parameters are
$r^2$ = 0.93 and $\sigma$ = 0.18 dex for the interval -2.44 $<$ [Fe/H] $<$
-0.08. Figure \ref{fig:q39-d14} shows the data points with
Eq. \ref{eq:q39-d14} plotted in the blue solid line, while the red dashed
line represents the curve fitted by C09 against their UVES
metallicities. Both curves agree well for [Fe/H] $\lesssim$ -0.4, but there
is a small divergence for the most metal-rich clusters. The C09 red curve
has higher metallicities for the most metal-rich clusters because they
adopted the higher metallicities for NGC~6528 and NGC~6553 from
\cite{carretta+01}, as discussed in the previous section. 

\subsection{Rutledge scale}
\label{rutscale}

\cite{rutledge+97} published a metallicity scale based on the reduced
equivalent widths ($W'$) of the near infrared CaII triplet lines for 52
clusters. We have 18 clusters in common and the best-fit quadratic function
relating their $W'$ values to our [Fe/H] determinations is given by 

\begin{equation}
\begin{array}{l}
{\rm [Fe/H]}_{\rm FORS2} = -2.65(\pm 0.28) + 0.13(\pm 0.17)\cdot
<W'_{\rm R97}>
+\\
+ 0.067(\pm 0.025)\cdot <W'_{\rm R97}>^2
\end{array}
\label{eq:r97-d14}
\end{equation}

\begin{figure}[!htb]
\centering
\includegraphics[width=\columnwidth]{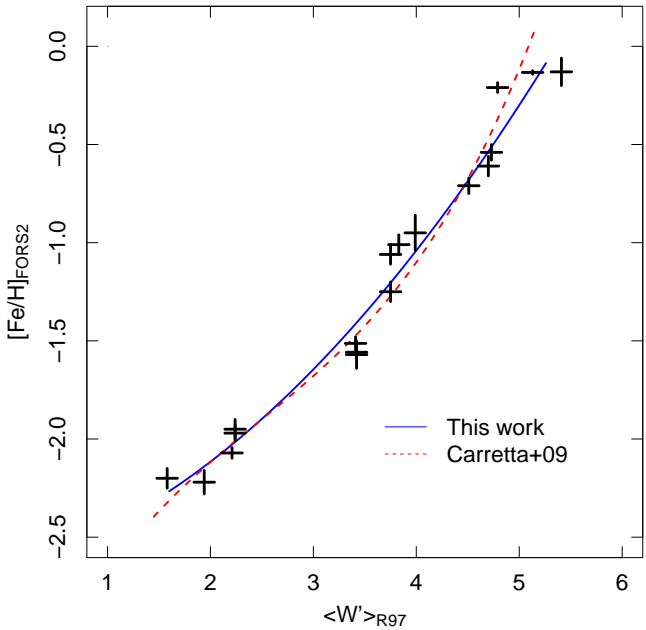}
\caption{Reduced equivalent width $<W'>$ from CaII triplet from
  \cite{rutledge+97} against [Fe/H] from this work. The blue
  solid line is the quadratic function fitted to 
  the data (Eq. \ref{eq:r97-d14}) and the red dashed line is the cubic
  function fitted by C09 to calibrate $<W'>_{\rm
    R97}$ to their scale. Fitted curves are shown only in their
  respective valid ranges.} 
\label{fig:r97-d14}
\end{figure}

\noindent The fit parameters are $r^2$ = 0.97
and $\sigma$ = 0.13 dex for the interval -2.27 $<$ [Fe/H] $<$
-0.08. Figure \ref{fig:r97-d14} displays the fitted curve as the blue solid
line, while the cubic function fitted by C09 is shown as the
dashed line. As in the case of Zinn \& West scale, Figure
\ref{fig:r97-d14} shows that our curve agrees well with that of
C09, with a slight discrepancy for clusters with [Fe/H] $\gtrsim$
-0.4, where the C09 relation gives higher metallicities for the metal-rich
clusters. The origin of this difference is as discussed above.

\subsection{Kraft \& Ivans scale}
\cite{kraft+03} collected a non-homogeneous set of high-resolution
stellar spectra of 16 clusters with [Fe/H] $<$ -0.7 and proceeded with
a homogeneous analysis. 
We have ten clusters in common with the Kraft \& Ivans abundances
related to ours by the
linear function given in Eq. \ref{eq:ki03-d14}:

\begin{equation}
\begin{array}{l}
{\rm [Fe/H]}_{\rm FORS2} = -0.16(\pm 0.12) + 0.94(\pm 0.08)\cdot {\rm [Fe/H]_{KI03}}
\end{array}
\label{eq:ki03-d14}
.\end{equation}

\begin{figure}[!htb]
\centering
\includegraphics[width=\columnwidth]{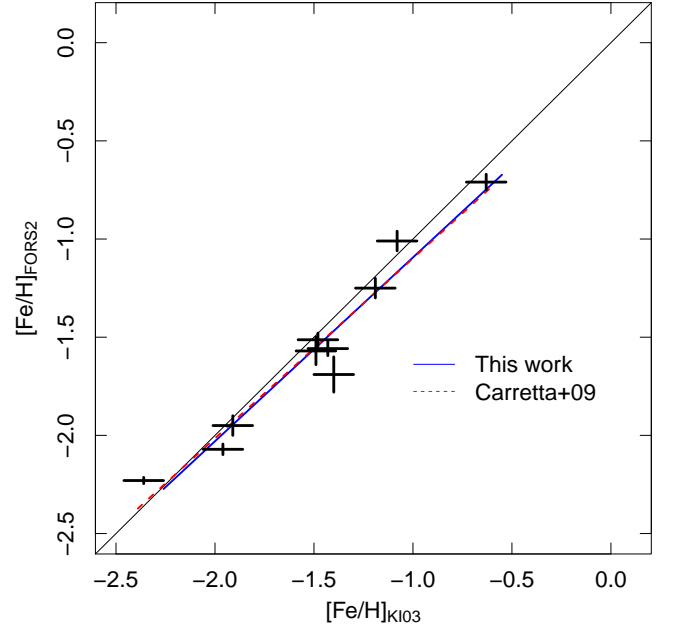}
\caption{[Fe/H] from \cite{kraft+03} against [Fe/H] from
  this work. The blue solid line is the linear function fitted to
  the data (Eq. \ref{eq:ki03-d14}) and the red dashed line is the linear
  function fitted by C09 to calibrate [Fe/H]$_{\rm
    KI03}$ to their scale. Fitted curves are shown only in their
  respective valid ranges.}
\label{fig:ki03-d14}
\end{figure}

\noindent The fit parameters are $r^2$ = 0.94
and $\sigma$ = 0.11 dex for the interval -2.28 $<$ [Fe/H] $<$
-0.66. Our relation, the blue line in Figure \ref{fig:ki03-d14}, and that of C09, the red dashed
line in the same figure, are essentially identical to the Kraft \& Ivans sample, which lacks clusters more
metal-rich than [Fe/H] $>$ -0.7 \citep{kraft+03}.

\subsection{Saviane scale}
\label{savscale}

\cite{saviane+12} analysed spectra from FORS2/VLT obtained in the same
project as the data presented here, but they analysed the CaII triplet
lines in a similar way to \cite{rutledge+97}. \cite{saviane+12} studied a total of
34 clusters, of which 14 were used as calibration clusters, and the other
20 were programme clusters. There are 27 clusters in
common and Eq. \ref{eq:s12-d14} shows the quadratic relation
between the  $<W'_{\rm S12}>$ values and our metallicities. The fit
parameters are $r^2$ = 0.97
and $\sigma$ = 0.12 dex for the interval -2.28 $<$ [Fe/H] $<$
-0.08: 

\begin{equation}
\begin{array}{l}
{\rm [Fe/H]}_{\rm FORS2} = -2.55(\pm 0.25) + 0.03(\pm 0.14)\cdot <W'_{\rm S12}> +\\
+ 0.068(\pm 0.018)\cdot <W'_{\rm S12}>^2
\end{array}
\label{eq:s12-d14}
\end{equation}

\begin{figure}[!htb]
\centering
\includegraphics[width=\columnwidth]{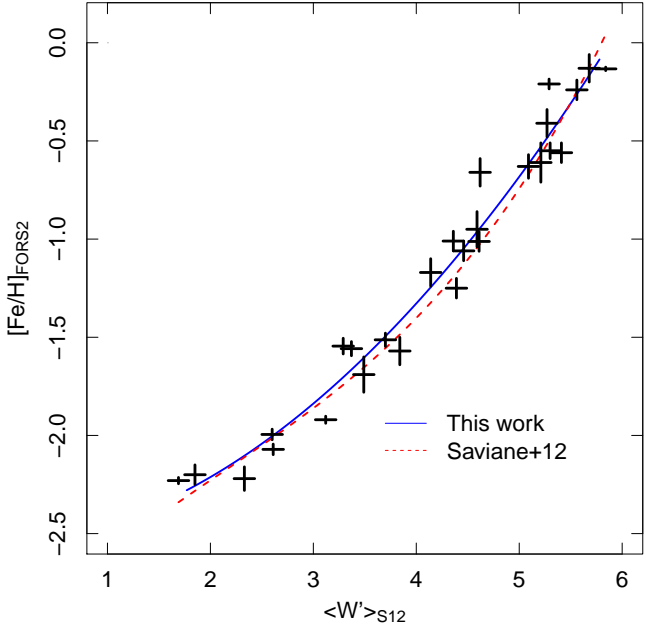}
\caption{Reduced equivalent width $<W'>$ from CaII triplet from
  \cite{saviane+12} against calibrated [Fe/H] from this work. The blue
  solid line is the quadratic function fitted to the data
  (Eq. \ref{eq:s12-d14}) and the red dashed line is the cubic function fitted
  by \cite{saviane+12} to calibrate their $<W'>_{\rm  S12}$ to the
  Carretta scale. Fitted curves are shown only in their
  respective valid ranges.}
\label{fig:s12-d14}
\end{figure}

\noindent Figure \ref{fig:s12-d14} shows the fit as the blue solid
line while the red dashed curve shows the calibration relation adopted by
 \cite{saviane+12}, which uses  the
metallicities from C09 as reference values. 
\cite{saviane+12} used metal-poor stars ([Fe/H] $<$ -2.5) to
conclude that their metallicity--line strength relation cannot be extrapolated, i.e.
it is only valid in the interval from $<W'_{\rm S12}> = 1.69$ to
$<W'_{\rm S12}> = 5.84$. The excellent agreement between the curves 
in Fig.\ \ref{fig:s12-d14} allows us to 
conclude that CaII triplet metallicities from FORS2/VLT spectra can be calibrated using 
metallicities for the same objects derived from visible spectra observed with the same instrument. 

%------------------------------------------------
\subsection{Conclusions on metallicity scales}

The information discussed in the preceding sections is compiled in Table \ref{tab:car09lit}. 
The metallicity range  is roughly the same for all the scales  with exception
of the \cite{kraft+03} scale, which does not have clusters
more metal-rich than [Fe/H] $\gtrsim$ -0.5. The largest homogeneous
sample is still that of \cite{zinn+84}, but their study is based on
integrated light which brings in a number of difficulties as discussed in the \cite{zinn+84}
paper. All the other data sets are based on measurements for
individual stars.  The largest homogeneous sample is then
that of \cite{rutledge+97}.  However, it is based on a CaII triplet index
which requires calibration to a [Fe/H] scale. C09 is
the largest sample based on high-resolution spectroscopy, although it has only 19
clusters with no cluster having [Fe/H] $>$ -0.4.
Our results from R $\sim$ 2,000 stellar spectra cover the entire
metallicity range of -2.4 $<$ [Fe/H] $<$ 0.0, and they are shown above to be
compatible with the high-resolution metallicities from C09, complemented
by metal-rich clusters from other high-resolution spectroscopic
studies.

\begin{table*}[!htb]
\caption{Summary of the properties of the current and previous metallicity scales. The fit
  parameters of previous metallicity scales against ours are presented. We  
  also give the characteristics of our metallicity scale for comparison.} 
\label{tab:car09lit}
\centering
\begin{tabular}{l@{ }c@{ }c@{ }c@{   }c@{   }c@{ }c@{ }c@{ }c@{ }c@{ }c@{ }c@{ }c}
\hline\hline
\noalign{\smallskip}
Scale                     & Total      & Avg. stars      &  $\lambda\lambda$& R & [Fe/H] &  Common   & $\sigma$  & r$^2$   & Polyn.  \\
                             & clusters   & per cluster    &  (nm)                       &    &        range     &  clusters     &                 &      &  order \\
\noalign{\smallskip}
\hline
\noalign{\smallskip}
this work      & 51                & 16              & 456 - 586 & 2,000 & [-2.4,  0.0]   & --- & ---              & ---  & --- \\
\noalign{\smallskip}
\hline
\noalign{\smallskip}
\cite{carretta+09} & 19                & 100            & 560 - 680 & 20,000 - 40,000 & [-2.4, -0.4]   & 13 & 0.07              & 0.99  & 1 \\
\cite{zinn+84}      & 56$^{(1)}$     & ---$^{(1)}$ & 360 - 570 & 775    & [-2.4, -0.1]  & 19 & 0.18 & 0.93 & 2 \\
\cite{rutledge+97}  & 52             &  19            &  725 - 900  &  2,000  &  [-2.3, -0.1]  & 17  & 0.13 & 0.97 & 2\\
\cite{kraft+03}      & 11+5$^{(2)}$ &     13             &  614 - 652  & 45,000 - 60,000         &  [-2.3, -0.7]  & 10  & 0.11       & 0.94 & 1 \\
\cite{saviane+12}  & 14+20$^{(3)}$        &  19           & 770 - 950  & 2,440  & [-2.3, -0.1]   &  14  & 0.12 & 0.97 & 2 \\
\noalign{\smallskip}
\hline
\end{tabular}
\tablefoot{ 
\tablefoottext{1}{They observed integrated spectra of 60 clusters; however, their Table 5 only presents metallicities for 56 objects.}
\tablefoottext{2}{They analysed spectra of different sources.}
\tablefoottext{3}{Observations and analysis are homogeneous, and 14 clusters were used for calibration.}
}
\end{table*}

\begin{figure}[!htb]
\centering
\includegraphics[width=\columnwidth]{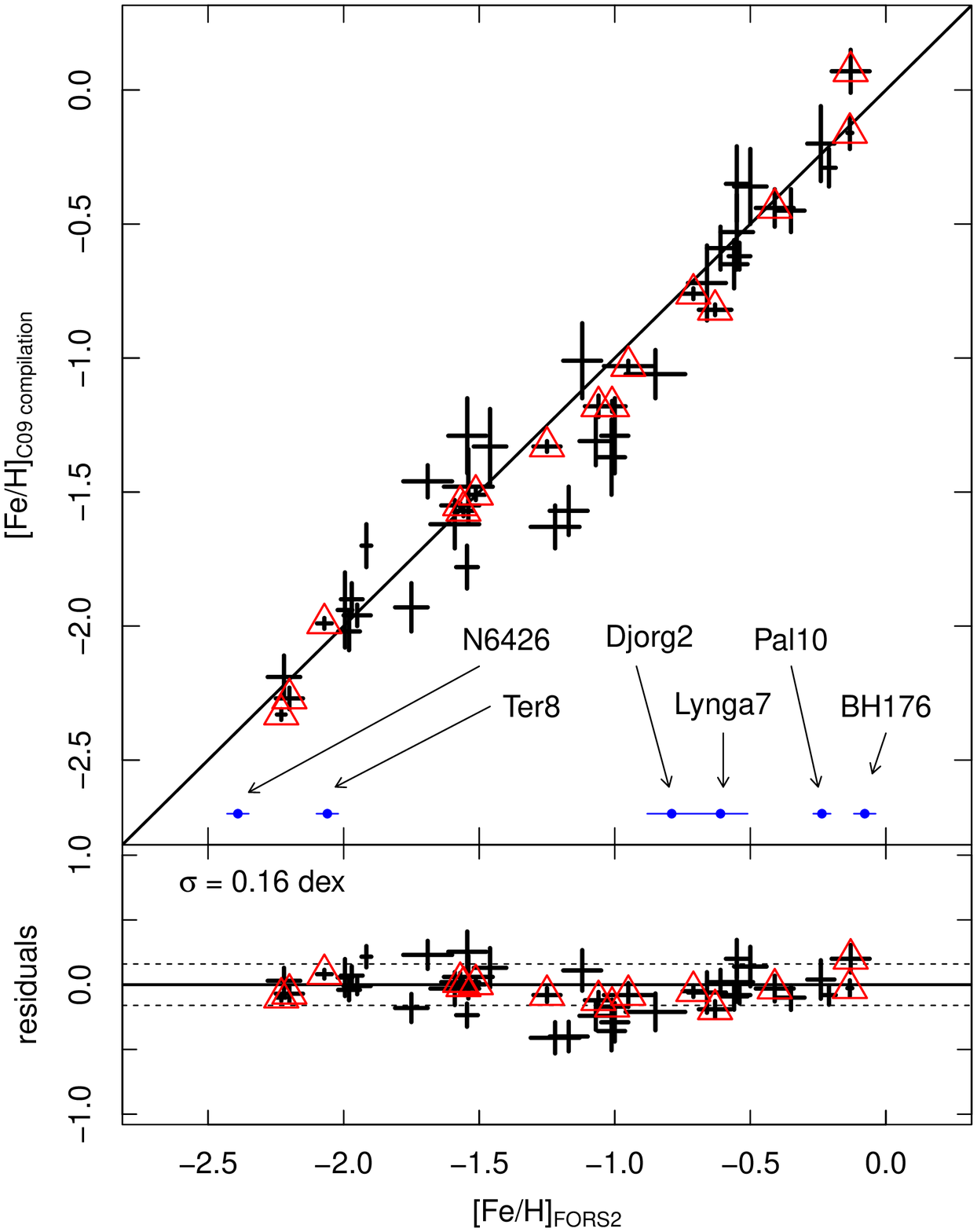}
\caption[Comparison of calibrated ${\rm [Fe/H]}$ with those from Carretta et
al. (2009) for the 51  clusters]{Comparison of the [Fe/H] values from
  this work with those from C09 for the 45
  clusters of our survey using values of Table
  \ref{finalparamavgcalib}. Red triangles emphasize the 15 clusters
  used to compare with C09 in Section \ref{feshcarscale}. Blue points are
  the six clusters not averaged by \cite{carretta+09} and analysed for the first time in a homogenous way
  in this work. A one-to-one line is plotted for reference. The
  residuals of the comparison are displayed in the bottom panel and
  have a standard deviation equal to 0.16~dex. 
}
\label{d14xc09-51}
\end{figure}

C09 calibrated all previous metallicity
scales to theirs and averaged them in order to get the best
metallicity estimate for all catalogued clusters. In Figure \ref{d14xc09-51}
we compare these values to those derived here for the Milky Way clusters in our FORS2 survey (see Table \ref{finalparamavgcalib}).  There are 45 clusters in common. 
A one-to-one line is plotted to guide the eye.  The [Fe/H] values are in
good agreement with the residuals shown in the bottom panel and reveal no trends with
abundance.  The 15 clusters used to compare our [Fe/H] determinations to the C09 
scale (cf.\ Sect. \ref{feshcarscale}) are highlighted as red triangles.
The dispersion of the residuals is $\sigma = 0.16$~dex, which is of
the order of the dispersion of the fit of all previous metallicity
scales to ours (Table \ref{tab:car09lit}). If C09 had averaged the
metallicity scales without having calibrating them to their scale, this
dispersion would be higher. Furthermore, the residuals do not show
trends with abundance, which supports the agreement of our results with C09 as
discussed in Sect. \ref{feshcarscale}.
Consequently, our metallicities are sufficiently robust to be used as references, from
the most metal-poor to solar-metallicity Milky Way globular clusters with a precision of $\sim$0.1~dex.
Moreover our data complement
the existing spectroscopic information on the Galactic GC system by
reducing the existing bias in the GC data base against distant and
reddened clusters.

We have also shown that
CaII triplet indices based on spectra from the same
instrumentation set-up can be calibrated using our [Fe/H] values, or C09's, producing
very similar results. Moreover this work also provides the
largest sample of homogeneous [Mg/Fe] and [$\alpha$/Fe] values for Milky Way
globular clusters. In addition, six clusters not contained in
\cite{carretta+09} have their metallicities determined 
from individual star spectra and a homogenous analysis for the first time. The clusters are
BH~176, Djorg~2, Pal~10, NGC~6426, Lynga~7, and Terzan~8 and they are shown as blue points in Fig. \ref{d14xc09-51}. Moreover, the first three clusters only had photometric metallicities estimations until now, and the available metallicity for NGC~6426 came from integrated spectroscopy and photometry only.

%------------------------------------------------
\section{Chemical evolution of the Milky Way}
\label{chem-MW}

The ratio [$\alpha$/Fe] plotted against
[Fe/H] provides an indication of the star formation efficiency
in the early Galaxy. Nucleosynthetic products from type II supernovae
(SNII) are effectively ejected shortly after the formation of
the progenitor massive star, releasing predominantly $\alpha$-elements 
together with
some iron\footnote{The ratio of [$\alpha$/Fe] released by SNII
  depends on the initial mass function. A typical value in the Milky
  Way is 0.4~dex \citep[e.g.][]{venn+04}.} into the interstellar medium.   
Type Ia supernovae (SNIa) of a given population, on the other hand, start to become
important from 0.3~Gyr to 3~Gyr after the SNII events, depending on
the galaxy properties \citep{greggio05}. These SN generate most of the Fe in
the Galaxy, decreasing the [$\alpha$/Fe] ratio. Magnesium is
one of the $\alpha$-elements and represents  these
processes well. Increasing values of [Fe/H] indicate subsequent
generations of stars so that lower metallicities and higher [$\alpha$/Fe]
stand for first stars enriched by SNII and higher 
metallicities and lower [$\alpha$/Fe] stand for
younger objects enriched by SNIa. The location of the turnover, designated by
[Fe/H]$_{\rm knee}$, identifies  when SNIa start to become
important. 

\begin{figure}[!htb]
\centering
\includegraphics[width=\columnwidth]{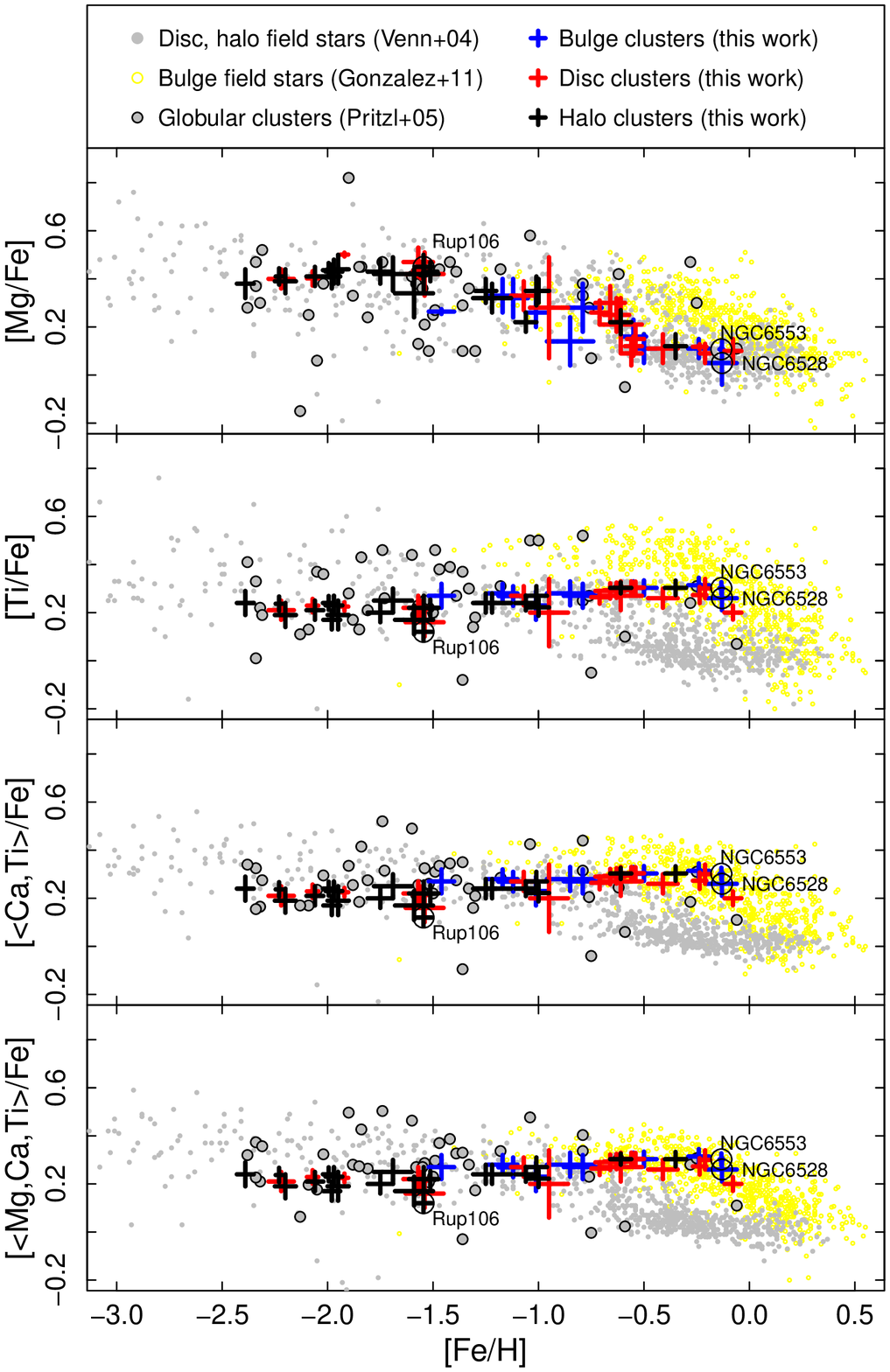}
\caption{[Mg/Fe] and [$\alpha$/Fe] for the 51 clusters from this work in
  comparison with disc and halo field stars from \cite{venn+04},
  bulge field stars from \cite{gonzalez+11}, and clusters from \cite{pritzl+05}.
  In the panels with [Ti/Fe], [$<$Ca,Ti$>$/Fe], and [$<$Mg,Ca,Ti$>$/Fe], our results are [$\alpha$/Fe] (see text for details). } 
\label{mg+alpha}
\end{figure}

Figure \ref{mg+alpha} displays the distribution along [Fe/H] of [Mg/Fe], [Ti/Fe],  two alternatives to represent the average [$\alpha$/Fe] (i.e. [$<$Ca,Ti$>$/Fe] and [$<$Mg,Ca,Ti$>$/Fe]) for halo and disc field stars from \cite{venn+04}, bulge field stars from \cite{gonzalez+11}, and clusters from \cite{pritzl+05}. 
 We overplot our results on [Mg/Fe] versus [Fe/H] in the uppermost panel, and [$\alpha$/Fe] versus [Fe/H] in the other three panels for the 51 globular clusters in our sample.  
The dispersion of our points is smaller than that of the Pritzl et al. points in all the panels. We note that our results were derived from homogeneous observations and analysis of R$\sim$2,000 spectra, while those from Pritzl et al. come from a compilation of different works based on higher resolution spectroscopy from the literature.

Whether globular clusters should follow the same pattern as field populations
or not is an open question. 
Qualitative analysis of the metal-poor region of the panels in Fig. \ref{mg+alpha} with [Fe/H] $<$ -1.0 shows that our results for [Mg/Fe] agree well with the Pritzl et al. clusters and also with disc+halo stars. Our results for [$\alpha$/Fe] reveal a positive slope, which leads to lower values with respect to [Ti/Fe], [$<$Ca,Ti$>$/Fe], and [$<$Mg,Ca,Ti$>$/Fe] for disc+halo stars. Nevertheless, the Pritzl et al. results also present a positive slope for [Ti/Fe] distribution, despite their large dispersion. It appears that  our results for [$\alpha$/Fe] are closer to those from Pritzl et al. for [Ti/Fe] than to the average of alpha-element enhancements.

Pritzl et al. do not have many clusters in the metal-rich regime where [Fe/H] $>$ -1.0 bulge stars clearly split from disc+halo stars and the [Fe/H]$_{\rm knee}$ is less obvious than that in the [Mg/Fe] panel; therefore, any comparison with their results would be poor. Our distribution of [Mg/Fe] follows that of disc+halo stars, while our [$\alpha$/Fe] is as enhanced as that for bulge stars.

\cite{pritzl+05} found a few peculiar
cases. Some of these are in common with our FORS2 survey data: two metal-rich
bulge clusters, NGC~6553 and NGC~6528, and the
metal-poor [$\alpha$/Fe]-depleted halo cluster, Rup~106. 
We indicate these clusters explicitly in Fig. \ref{mg+alpha}; in
  particular, the lower three panels with [$\alpha$/Fe] confirm that Rup~106 has a lower
  [$\alpha$/Fe] ratio than the  other clusters and also lower than halo and
  disc stars at similar metallicities.
The bulge clusters NGC~6553 and NGC~6528 follow the bulge stars.
We have shown in Paper I that our
abundances of [$\alpha$/Fe] for NGC~6528 and NGC~6553 are in
agreement with high-resolution spectroscopic results.
We were able to recover a subtle depletion in [$\alpha$/Fe] for Rup~106 and an enhancement in [$\alpha$/Fe] for NGC~6528 and NGC~6553.

We note that our [$\alpha$/Fe] is derived from the comparison with the Coelho library; here  the spectra are modelled by varying all $\alpha$-elements O, Mg, S, Si, Ca, and Ti. The [$\alpha$/Fe] distribution matches that of [Ti/Fe] better than other elements and does not show the turnover. It is interesting to note that in the metal-rich regime, \cite{lecureur+07} found enhancements of Na and Al; therefore, the  metal-rich bulge stars might show other unexpected behaviour. Further checks are underway that vary each element individually, rather than varying all alpha-elements together
as is done in \cite{coelho+05}.
The analysis approach could be improved in the future  by including
stars from the Magellanic Clouds, which  generally have lower [$\alpha$/Fe] than  the
Galaxy at higher metallicities \citep[e.g.][]{vanderswaelmen+13}.

%
%________________________________________________________________
\section{Horizontal branch morphology and the second parameter problem}
\label{HB-sect}

\begin{figure}[!htb]
  \centering
 \includegraphics[width=\columnwidth]{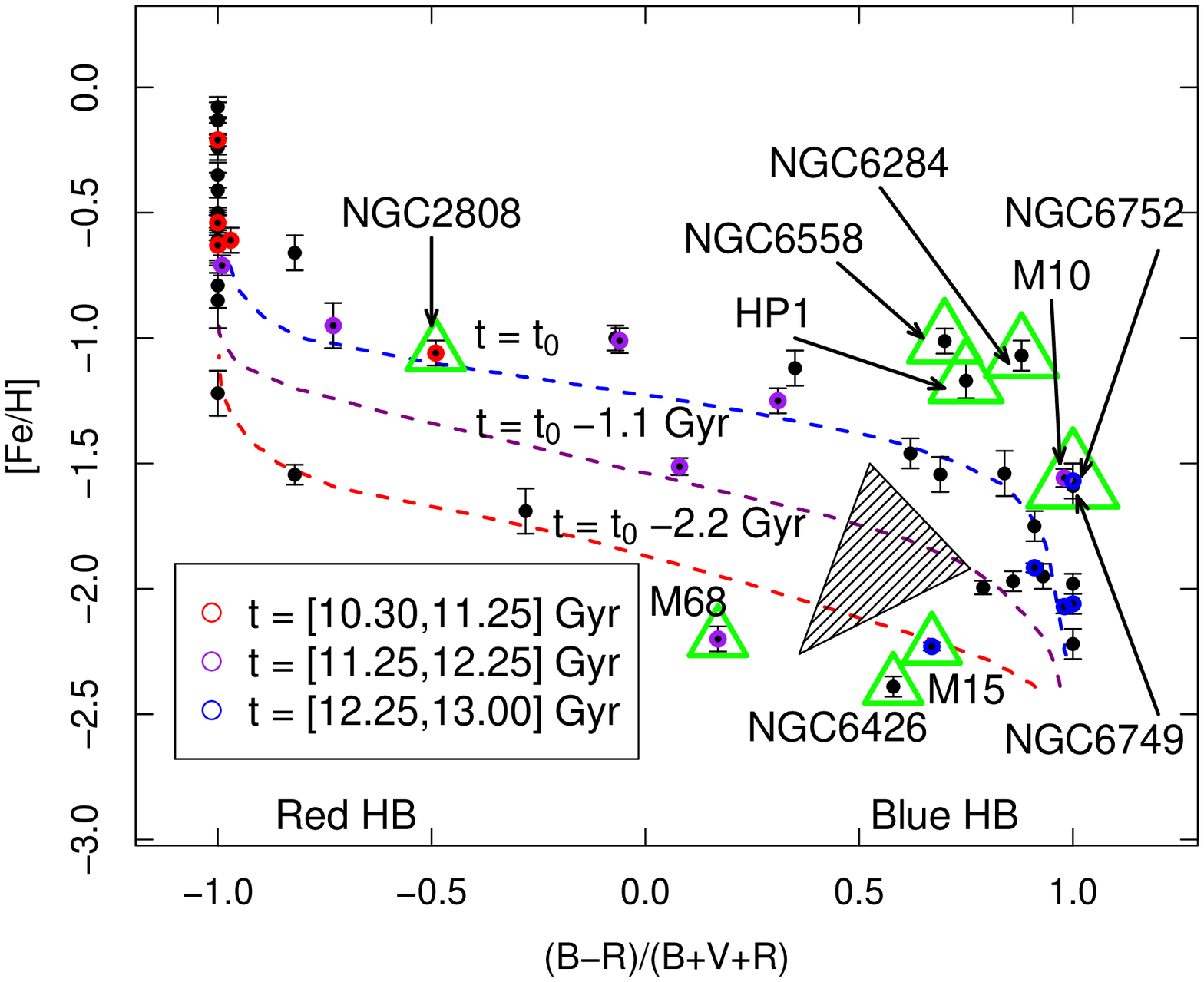}
  \caption{Metallicity as a function of horizontal branch morphology
    (HB index) for all 51 clusters of our sample, where HB index is
    from \cite{mackey+05}. Three isochrones from \cite{rey+01} are
    overplotted, where t$_0$ is the mean age of inner halo
      clusters as defined by Rey et al. as R$_{\rm GC} < 8$~kpc. Ages
    from \cite{vandenberg+13} are available only for 
  17 out of 51 clusters in our sample, and are shown as red, purple,
  and blue circles for young, intermediate-age, and old clusters. The
  hatched triangle shows the region of the Oosterhoff gap as defined by
  \cite{catelan09}.}
  \label{hbmorpho}
\end{figure}

The horizontal branch (HB) morphology in a CMD of a
globular cluster is shaped mainly by metallicity, but there are other
parameters that influence its predominant colour. 
These include age, helium abundance,
CNO abundance, and RGB mass-loss, among others (see review of
\citealp{catelan09} and references therein). All phenomena
may be shaping the HB together, with one more important than the
others; for example, the second parameter is traditionally assumed to be age, 
but there are exceptions \cite[e.g.][]{fusi-pecci+97}.
Figure \ref{hbmorpho} shows the effect of age (from
\citealp{vandenberg+13} when available) and metallicity (from this work)
on the colour of the horizontal branch, with the HB index being
(B-R)/(B+V+R), where B, R, and V are the number of blue, red, and
variable stars \citep{lee+94}. The older or more metal-poor the cluster, the bluer
the HB; redder HB represents younger and/or more metal-rich
clusters. 
Three HB isochrones with different ages from \cite{rey+01} are shown.

We call attention to four groups of clusters in
the plot, all of them indicated in Fig. \ref{hbmorpho}:

\begin{itemize}
\item{NGC~2808: typical bimodal HB \citep[e.g.][]{corwin+04}.}
\item{M~68, NGC~6426, and M~15: M 68 possibly has age as the
    second parameter \citep{vandenberg+13}. NGC~6426 is older,
    contrary to what is expected from Fig. \ref{hbmorpho}
    \citep{hatzidimitriou+99}. M~15 is one of the two clusters
    (of 16) that do not follow the blue HB distribution of field
    stars \citep{brown+05}.}
\item{M~10, NGC~6752, and NGC~6749: the first has a HB morphology
    possibly justified by He variations \citep{gratton+10}. The second
  has a very complex HB morphology \citep{momany+02}. The third
  cluster has a CMD from \cite{rosino+97} and no discussion on the second
  parameter problem.}
\item{HP~1, NGC~6558, and NGC~6284: the first two have been studied by
  \citep{ortolani+97,ortolani+11,barbuy+06,barbuy+07} and are candidates for  the oldest clusters in the bulge. The third has a CMD from HST
  observations and no discussion about HB morphology \citep{piotto+02}.}
\end{itemize}

The first three groups have been well discussed in the literature, with
  the exception of NGC~6749, which  should be analysed in more detail.
  For a review on the topic we refer to
  \cite{catelan09} and \cite{gratton+10}. To look for the oldest clusters in the
  Milky Way, we focus the discussion of the last group as follows.

For this group of clusters with blue HB and [Fe/H]$\sim$-1.0,
there are some possible explanations: (i) the clusters 
are older than all others; (ii) their He abundance is lower; or (iii)
their CNO abundance is lower. Figure \ref{hbmorpho} can only reveal if age
and metallicity are able to explain the HB morphology. Fixing the HB
index and [Fe/H] and varying only age, these clusters would be the
oldest objects in the Milky Way by projecting the age gradient from the
isochrones upwards in the plot. As a result, deriving ages and HB star
abundances for these clusters is crucial to make such strong
conclusion. \cite{vandenberg+13} have not published ages for these
clusters, but other papers have, as is discussed below.

\paragraph{NGC~6284}

This is a disc cluster located behind the bulge with E(B-V) = 0.28,
7.5~kpc  from the Galactic centre and out of the projected plane
of the X-shaped bulge. This location likely rules out the possibility of NGC~6284
being a bulge cluster ejected by the dynamics of the
``X''. \cite{catelan09} noticed some peculiarities about 
NGC~6284 and classified it as an Oosterhoff-intermediate globular
cluster, yet it does not fall in the region (indicated by a
triangle in Figure \ref{hbmorpho}) where such clusters are expected. 
\cite{piotto+02} presented a HST-based CMD showing a clear blue HB,
therefore its index is verified. We derived a metallicity of [Fe/H] =
-1.07$\pm$0.06 for this cluster which is more metal-rich yet
still compatible to within 2.2-$\sigma$ with the value of -1.31$\pm$0.09 (Q39
index from \citealp{zinn80}  calibrated to the scale presented by
\citealp{carretta+09}). \cite{meissner+06} derived
11.00$\pm$0.25~Gyr\footnote{They do not provide an error bar, but
  their age resolution is 0.5~Gyr and we assume half of it as an
  estimate of internal error.} for NGC~6284, which is relatively young
for a globular cluster and may rule out the proposition that
NGC~6284 could be among the oldest objects in the Milky Way.
For this cluster, even if age is helping to shape the
blue HB, a lower He and/or CNO abundance should be important
factors.

\paragraph{HP~1}

This bulge cluster is the innermost globular cluster known in the
Milky Way, with E(B-V) = 1.12 and only 500~pc  from the Galactic
centre where Sgr A* with the central black hole and surrounding
nuclear star cluster are located \citep{genzel+10}. The contamination
of foreground and background stars and dust is very high, and
\cite{ortolani+11} performed a decontamination using proper
motion with a baseline of 14 years. Even with 
Multi-Conjugate Adaptive Optics at the VLT (MAD/VLT) 
photometry producing a well-defined CMD, it only reaches the
subgiant branch and the main sequence turnoff is undersampled. With
this information they estimated an age of 13.7~Gyr 
relative to other well-studied clusters. The respective isochrone
(assuming Z=0.002, [Fe/H] $\approx$ -0.9) agrees well with the current
CMD. A lower limit for the age for HP~1 based on their method
would be 12.7~Gyr. This result supports the prediction of an old age from Figure
\ref{hbmorpho}. We derived [Fe/H] = -1.17$\pm$0.07 from the average of eight
red giant stars in the cluster, which is compatible with [Fe/H] =
-1.0$\pm$0.2 found by \cite{barbuy+06} 
from the analysis of high-resolution UVES spectra of two red giant
stars. They derived [Mg/Fe] = 0.10 and we found a more alpha-enhanced
ratio comparable to bulge field stars of similar metallicity, [Mg/Fe]
= 0.33$\pm$0.07.
We confirm that HP~1 is one of the top candidates for  
the oldest globular cluster in our Galaxy, sharing the age of the
Milky Way. 
The orbit of HP~1 was derived by \cite{ortolani+11} and Rossi et al. (2016, in preparation),
showing that it is confined within the bulge/bar.
The central region was the  densest environment of the
proto-galaxy where  globular clusters probably formed first.
 Deeper photometry is
needed to better sample the main sequence turnoff and to have a
definitive isochrone fitting, which makes it a perfect target for  ACS/HST or the
forthcoming E-ELT.

\paragraph{NGC~6558}

This bulge cluster was extensively discussed in Paper I where we show
the compatibility of our results with those of \cite{barbuy+07}, star
by star. Therefore, we concentrate  on further discussion about its
role in this special group in the [Fe/H]-HB index plot. We also
highlight the new abundance uncertainty using the updated criteria
described in Sect. \ref{sec:met}: [Fe/H] =  -1.01$\pm$0.05 and [Mg/Fe] =
0.26$\pm$0.06. \cite{barbuy+07} derived [Fe/H] = -0.97$\pm$0.15 and
[Mg/Fe] = 0.24, which are compatible with our results. The horizontal branch
is very similar to that of HP~1 \citep{barbuy+07,ortolani+11};
therefore, the position of NGC~6558 in Fig. \ref{hbmorpho} is valid.
The age of the cluster is a more difficult
matter. \cite{barbuy+07} have fitted two isochrones of 14~Gyr in a
CMD containing cluster and field stars; \cite{alonso-garcia+12} has
shown a differential reddening varying from -0.06 to +0.08 with
respect to the average E(B-V) = 0.44; and 
\cite{rossi+15} have published a proper motion cleaned CMD which shows a
broad red giant branch but with less deep photometry and an
undersampled main sequence turnoff.
These complexities may lead to
uncertainties in the age derivation, but the spread of the main
sequence turnoff is less than $\Delta$V $\approx$ 0.2mag, which would make it difficult to measure the relative age to better than 1 Gyr.
We conclude that NGC~6558 should
not be classified among the younger globular clusters. Consequently, it may be
that age is a strong candidate for the second parameter in the case of
this cluster
causing a blue HB and placing it as one of the oldest objects in the
Milky Way. As proposed for HP~1 above, high-resolution spectroscopy of
HB stars is needed in order to understand the role of He and CNO and further constrain
the age.

%
%________________________________________________________________

\section{Summary and conclusions}
\label{sec:conc}

In this work we present parameters -- derived from R$\sim$2,000 visible spectra by applying the 
methods described in Paper I -- for 51 Galactic globular clusters.  We observed 819 red
giant stars and analysed 758 useful spectra; of these we classified 464 stars as members
of the 51 clusters and 294 as non-members.  Membership 
selection included deriving radial velocities for all 758 spectra. Estimates for T$_{\rm
  eff}$, log($g$), and [Fe/H]  were determined by using observed (MILES) and
   synthetic (Coelho) spectral libraries and the results from both libraries
averaged for the final results. We compared our results with
six previous metallicity scales and fit polynomial functions with
coefficients of determination $r^2 \geq 0.93$ and $\sigma \leq 0.18$~dex. The
most important comparison is against C09, which
contains the largest sample of clusters (19) with abundances based on high-resolution
spectroscopy. For this case, a linear fit was very good
with $r^2$ = 0.99 and $\sigma$ = 0.08~dex. The slope of the fit  is compatible with 1.0 and the offset is near zero, which means that our metallicity results are in excellent agreement with those from high-resolution spectroscopy in the range -2.5 $\lesssim$ [Fe/H] $\lesssim$ 0.0 with no need to apply any scale or calibration.
The other scales are based on lower resolution spectroscopy,
CaII triplet, limited sample, or integrated light, and the functions
fitted against our metallicities are compatible with those fitted
against the C09 results, except for the metal-rich regime for which we used updated and robust references from  high-resolution spectroscopy. Metal-rich clusters with [Fe/H] $\gtrsim$ -0.5 are less metal-rich than the findings of C09.
An important consequence of our results is that
CaII triplet line strengths, such as those of \cite{saviane+12}, can be calibrated directly by applying our
approach to visible region spectra of the same stars obtained with the same instrument.

C09 took an average of metallicities available
at that time for all globular clusters from different metallicity
scales after calibration to their scale.  For the 45 clusters in common with our
sample, the comparison has no trends with abundance and a dispersion of $\sigma$ =
0.16~dex, in agreement with our comparison to the same metallicity scales.
The metallicities derived in this work are
robust to within 0.1~dex for the entire range of [Fe/H] shown by Galactic
globular clusters. Six clusters of our sample do not have previous measurements presented in
  Carretta's scale.  The clusters are BH~176, Djorg~2, Pal~10, NGC~6426, Lynga~7, and Terzan~8 and we present abundances for these clusters in a homogeneous scale for
  the first time. Moreover, the first three clusters have only had photometric metallicities estimations until now, and the available metallicity for NGC~6426 came only from integrated spectroscopy and photometry.
  
 Another important product of this survey is that we also provide
[Mg/Fe] and [$\alpha$/Fe] for all 758 stars and the average values for member
stars in the 51 clusters on a homogeneous scale. This is the largest
sample of $\alpha$-element abundances for Milky Way globular clusters
using the same set-up for observations and same method of analysis.
The distribution of [Mg/Fe] with [Fe/H] for the 51 clusters follows the
same trends as for field stars from the halo and disc, but does not recover
the peculiar $\alpha$-element depletion for the metal-poor halo cluster
Rup~106, and does not support high [$\alpha$/Fe] for clusters like NGC~6553 and NGC~6528.
The [$\alpha$/Fe], [Fe/H] relation follows the trend of bulge stars, 
and recovers abundances for NGC~6553, NGC~6528 compatible with bulge field
stars, as well as the depletion in [$\alpha$/Fe] for Rup~106. 
However, the distributions of [Mg/Fe] and [$\alpha$/Fe] with [Fe/H] do not agree
well with each other  possibly because [$\alpha$/Fe] is derived from the
comparison with the Coelho library, which models the spectra by varying all
$\alpha$-elements O, Mg, S, Si, Ca, and Ti, while for the clusters the observed
[$\alpha$/Fe] is the average of [Mg/Fe], [Ca/Fe], and [Ti/Fe]
only. We intend to improve
$\alpha$-element abundance measurements in a future paper.

The metallicities derived in this work were plotted against the
index of horizontal branch morphology and we identified four
peculiar groups in the diagram. We then focused on the group containing the metal-rich
and blue horizontal branch clusters HP~1, NGC~6558, and NGC~6284.  These clusters are
candidates for the oldest objects in the Milky Way.
HP~1 and NGC~6558 possess bluer horizontal branch morphologies than
expected for their metallicities of [Fe/H] = -1.17$\pm$0.07 and
-1.01$\pm$0.05, respectively.  If the second parameter that drives the morphology of the
        horizontal branch in these clusters is age, then they are indeed likely to be very old
        objects.  This is consistent with previous work that has shown
that the two bulge clusters share the age of the Milky Way. NGC~6284
also has a blue horizontal branch and a relatively high metallicity of [Fe/H] = -1.07$\pm$0.06.
However, existing studies have shown that it is a few Gyr younger than the other clusters. 
Therefore, the second parameter for this cluster may not be age, but is perhaps related to CNO or He 
abundances.  Further studies are warranted.

%
%________________________________________________________________

\begin{acknowledgements}
BD acknowledges support from CNPq, CAPES, ESO, and the European Commission's
Framework Programme 7, through the Marie Curie International Research
Staff Exchange Scheme LACEGAL (PIRSES-GA-2010-269264). BD also
acknowledges his visit to EH at the Osservatorio di Padova and the
visit of GDC to ESO for useful discussions on this paper.
BB acknowledges partial financial support from CNPq, CAPES, and Fapesp.
Veronica Sommariva is acknowledged for helping with part of the data reduction.
The authors acknowledge the anonymous referee for the very useful comments and suggestions.
\end{acknowledgements}

%
%________________________________________________________________

\bibliographystyle{aa}
\bibliography{bibliography}

%
%________________________________________________________________

\longtab{1}{
\scriptsize
\begin{longtable}{llllccrlllc}
\caption{\label{logtable} Log of
  observations of the 51 globular clusters using FORS2/VLT with grism
  1400V. Classification of each cluster as
  (B)ulge, (D)isc, inner (IH) or outer halo (OH), as well as, open
  cluster (DOpen) and dwarf galaxy-related cluster (OHdSph) follow
  the criteria defined by \cite[][C10]{carretta+10}, except where indicated
  the contrary following the classification of \cite[][B15]{bica+15} for bulge clusters. The adopted classification is
  explicitly displayed. In the last column we show the 
 numbers of analysed stars that belong to each cluster (C) and those that we classified
 as field stars (F).}\\ 
\hline\hline
\noalign{\smallskip}
Cluster   & Other     & $\alpha$(J2000)                       & $\delta$(J2000)                       & obs. date & UT    & $\tau$  & Pop. (C10)  & Pop. (B15) & Pop. & \# stars\\
    & names           & $^{\rm h}$ $^{\rm m}$ $^{\rm s}$ & $^{\circ}$ ' '' & dd.mm.yyyy  & h:m:s &   (s)  & &  & (adopted) & C/F\\
\noalign{\smallskip}
\hline
\noalign{\smallskip}
\endfirsthead
\caption[]{continued.}\\
\hline\hline
\noalign{\smallskip}
Cluster   & Other     & $\alpha$(J2000)                       & $\delta$(J2000)                       & obs. date & UT    & $\tau$  & Pop. (C10)  & Pop. (B15) & Pop. & \# stars\\
    & names           & $^{\rm h}$ $^{\rm m}$ $^{\rm s}$ & $^{\circ}$ ' '' & dd.mm.yyyy  & h:m:s &   (s)  & &  & (adopted) & C/F\\
\noalign{\smallskip}
\hline
\noalign{\smallskip}
\endhead
\hline
\endfoot
\noalign{\smallskip} 
NGC 104$^a$              &  47 Tuc      &  00 24 05.67 & -72 04 52.6 &  22.10.2001  & 07:16:53  &  120.0  &    D       & --- &  D &   15/1   \\       
NGC 2298$^a$         &              &  06 48 59.41 & -36 00 19.1 &  23.10.2001  & 06:25:42  &  120.0  &    OH      & --- & OH &   5/0   \\         
NGC 2808$^d$         &              &  09 12 03.10 & -64 51 48.6 &  29.05.2006  & 00:06:07  &   45.0  &    IH      & --- &  IH & 14/4   \\         
NGC 3201$^d$         &              &  10 17 36.82 & -46 24 44.9 &  28.05.2006  & 22:52:56  &   20.8  &    IH      & --- &  IH & 13/2   \\         
NGC 4372$^c$         &              &  12 25 45.40 & -72 39 32.4 &  25.05.2003  & 01:34:41  &  300.0  &    D       & --- &  D &  8/2   \\          
Rup 106$^d$              &              &  12 38 40.2  & -51 09 01   &  28.05.2006  & 23:15:01  &  758.6  &    OH      & --- &   OH & 8/7   \\       
NGC 4590$^b$         &  M 68        &  12 39 27.98 & -26 44 38.6 &  07.05.2002  & 03:38:10  &   60.0  &    IH      & --- &  IH &  7/3   \\         
NGC 5634$^e$         &              &  14 29 37.23 & -05 58 35.1 &  26.06.2012  & 00:12:03  &  240.0  &    OH      & --- &   OH & 8/1   \\         
NGC 5694$^e$         &              &  14 39 36.29 & -26 32 20.2 &  25.06.2012  & 23:27:18  &  540.0  &    OH      & --- &   OH & 8/3   \\         
NGC 5824$^d$         &              &  15 03 58.63 & -33 04 05.6 &  29.05.2006  & 00:26:28  &  553.3  &    OH      & --- &   OH & 15/3   \\        
NGC 5897$^b$         &              &  15 17 24.50 & -21 00 37.0 &  07.05.2002  & 03:53:29  &   60.0  &    IH      & --- &   IH & 8/0   \\         
NGC 5904$^c$         &  M 5         &  15 18 33.22 & +02 04 51.7 &  04.05.2003  & 06:00:38  &  300.0  &    IH      & --- &  IH &  9/0   \\          
NGC 5927$^b$         &              &  15 28 00.69 & -50 40 22.9 &  07.05.2002  & 04:14:29  &  300.0  &    D       &  D  &    D & 6/0   \\         
NGC 5946$^e$         &              &  15 35 28.52 & -50 39 34.8 &  23.06.2012  & 02:52:00  &  180.0  &    IH      & --- &   IH & 5/10   \\        
BH 176$^e$                   &              &  15 39 07.45 & -50 03 09.8 &  22.05.2012  & 03:06:04  &  600.0  &    DOpen   &  D  &  D & 11/4   \\       
Lynga 7$^d$              &  BH 184      &  16 11 03.65 & -55 19 04.0 &  29.05.2006  & 01:17:18  &  451.6  &    D       &  D  &   D & 3/10   \\       
Pal 14$^e$                   &  AvdB        &  16 11 00.6  & +14 57 28   &  15.06.2012  & 02:49:24  & 1140.0  &    OH      & --- &   OH & 6/1   \\   
NGC 6121$^d$         &  M 4         &  16 23 35.22 & -26 31 32.7 &  29.05.2006  & 02:53:01  &    5.8  &    IH      & --- &    IH & 8/6   \\   
NGC 6171$^b$         &  M 107       &  16 32 31.86 & -13 03 13.6 &  07.05.2002  & 04:39:29  &   60.0  &    B       &  non-B &  D &  1/4*   \\              
NGC 6254$^d$         &  M 10        &  16 57 09.05 & -04 06 01.1 &  29.05.2006  & 03:11:21  &   54.5  &    D       & --- &   D & 13/2   \\         
NGC 6284$^e$         &              &  17 04 28.51 & -24 45 53.5 &  22.07.2012  & 03:17:03  &  180.0  &    D       & --- &  D & 7/10   \\          
NGC 6316$^e$         &              &  17 16 37.30 & -28 08 24.4 &  22.07.2012  & 03:29:22  &  180.0  &    B       &  B &   B & 7/9   \\           
NGC 6356$^d$         &              &  17 23 34.93 & -17 48 46.9 &  29.05.2006  & 04:40:47  &  167.9  &    D       & --- &  D & 13/5   \\          
NGC 6355$^e$         &              &  17 23 58.59 & -26 21 12.3 &  11.09.2012  & 23:19:52  &  240.0  &    B       &  B &   B & 6/10   \\          
NGC 6352$^e$         &              &  17 25 29.11 & -48 25 19.8 &  22.05.2012  & 05:33:04  &   60.0  &    B       &  non-B &   D & 12/2   \\              
NGC 6366$^e$         &              &  17 27 44.24 & -05 04 47.5 &  15.06.2012  & 04:53:54  &   60.0  &    IH      & --- &  IH & 14/3   \\         
HP 1$^d$                     &  BH 229      &  17 31 05.2  & -29 58 54   &  30.05.2006  & 05:19:08  & 1037.5  &    B       &  B &   B & 8/19   \\       
NGC 6401$^e$         &              &  17 38 36.60 & -23 54 34.2 &  14.07.2012  & 05:36:29  &  300.0  &    B       &  B &  B & 6/12   \\           
NGC 6397$^{c}$       &              &  17 40 42.09 & -53 40 27.6 &  06.05.2003  & 03:54:19  &  300.0  &    D       & --- &  D & 18/3   \\ 
NGC 6397$^{d}$       &              &  ''          & ''          &  29.05.2006  & 05:24:27  &   7.7   &    ''      & ''  &   ''  & ''    \\ 
Pal 6$^e$                &              &  17 43 42.2  & -26 13 21   &  12.09.2012  & 01:01:28  & 780.0   &    B       &  B &  B & 4/13   \\       
NGC 6426$^e$         &              &  17 44 54.65 & +03 10 12.5 &  13.07.2012  & 02:31:12  & 500.0   &    IH      & --- &  IH &  5/5   \\         
NGC 6440$^d$         &              &  17 48 52.70 & -20 21 36.9 &  20.05.2006  & 05:38:49  & 649.2   &    B       &  B &   B & 7/9   \\           
NGC 6441$^d$         &              &  17 50 13.06 & -37 03 05.2 &  29.05.2006  & 06:27:01  & 227.2   &    D       &  non-B &  D & 8/10   \\ 
NGC 6453$^e$         &              &  17 50 51.70 & -34 35 57.0 &  12.09.2012  & 00:50:14  & 300.0   &    D       & --- &   D & 3/13   \\                 
Djorg 2$^e$              & ESO456-SC38  &  18 01 49.1  & -27 49 33   &  14.07.2012  & 05:52:59  & 180.0   &    B       &  B &  B & 4/11   \\       
NGC 6528$^d$         &              &  ''          & ''          &  29.05.2006  & 08:36:22  & 149.4   &    B       &  B &   B & 4/13   \\    
NGC 6539$^e$         &              &  18 04 49.68 & -07 35 09.1 &  12.09.2012  & 00:13:01  & 360.0   &    B       & B  &   B & 7/8   \\           
NGC 6553$^d$         &              &  18 09 17.60 & -25 54 31.3 &  29.05.2006  & 08:57:50  &  79.4   &    B       &  B &  B & 11/6   \\           
NGC 6558$^d$         &              &  18 10 17.60 & -31 45 50.0 &  29.05.2006  & 06:55:32  & 148.3   &    B       &  B &   B & 4/13   \\          
IC 1276$^d$              &  Pal 7       &  18 10 44.20 & -07 12 27.4 &  29.05.2006  & 07:17:06  & 229.8   &    D       &  non-B &  D & 12/5   \\       
NGC 6569$^d$         &              &  18 13 38.80 & -31 49 36.8 &  29.05.2006  & 07:43:05  & 210.4   &    B       &  non-B &  D & 7/11   \\               
NGC 6656$^d$         &  M 22        &  18 36 23.94 & -23 54 17.1 &  29.05.2006  & 08:22:25  &   36.1  &    D       & --- &  D & 44/9   \\          
''                   &  ''          &  ''          & ''          &  ''          & 09:57:32  &   36.1  &    ''      & ''  &   '' & ''    \\ 
''                   &  ''          &  ''          & ''          &  ''          & 08:08:05  &   36.1  &    ''      & ''  &    '' & ''    \\ 
NGC 6749$^e$         &              &  19 05 15.3  & +01 54 03   &  27.05.2012  & 05:04:40  &  810.0  &    IH      & --- &  IH & 4/13   \\         
NGC 6752$^a$         &              &  19 10 52.11 & -59 59 04.4 &  25.05.2003  & 06:55:40  &  300.0  &    D       & --- &   D & 5/1   \\          
Pal 10$^e$                   &              &  19 18 02.1  & +18 34 18   &  16.06.2012  & 05:38:57  &  900.0  &    D       & --- &  D & 9/14   \\       
Terzan 8$^e$         &              &  19 41 44.41 & -33 59 58.1 &  12.07.2012  & 07:37:44  &  360.0  &    OHdSph  & --- &  OH & 12/1   \\
''                   &              &  ''          & ''          &  13.07.2012  & 08:10:51  &  360.0  &    ''      & ''  &   '' & ''    \\ 
''                   &              &  ''          & ''          &  14.07.2012  & 05:18:10  &  360.0  &    ''      & ''  &    '' & ''    \\ 
Pal 11$^e$                   &              &  19 45 14.4  & -08 00 26   &  13.06.2012  & 07:32:42  &  180.0  &    IH      &  non-B &   IH & 10/2   \\       
''                           &              &  ''          & ''          &  ''          & 07:50:37  &  300.0  &    ''      & ''  &   '' & ''    \\ 
NGC 6838$^d$         &  M 71        &  19 53 46.49 & +18 46 45.1 &  29.05.2006  & 09:14:32  &   17.2  &    D       &  D  &  D &  8/4   \\           
NGC 6864$^e$         &  M 75        &  20 06 04.69 & -21 55 16.2 &  27.07.2012  & 05:35:32  &  240.0  &    IH      & --- &   IH & 10/2   \\        
NGC 7006$^d$         &              &  21 01 29.38 & +16 11 14.4 &  30.05.2006  & 09:08:54  & 1200.0  &    OH      & --- &   OH & 5/9   \\         
NGC 7078$^d$         &  M 15        &  21 29 58.33 & +12 10 01.2 &  29.05.2006  & 09:30:56  &   47.7  &    IH      & --- &  IH & 15/0   \\         
\noalign{\smallskip}
\end{longtable}
\tablefoot{ 
\tablefoottext{a}{2001 observations, ID 68.B-0482(A).}
\tablefoottext{b}{2002 observations, ID 69.D-0455(A).}
\tablefoottext{c}{2003 observations, ID 71.D-0219(A).}
\tablefoottext{d}{2006 observations, ID 077.D-0775(A).}
\tablefoottext{e}{2012 observations, ID 089.D-0493(B).}
\tablefoottext{*}{Membership selection for M~107 was not very clear,
  but one star matches literature values of T$_{eff}$, log($g$),
  [Fe/H] and we considered that as member.} 
}
\normalsize
}

\longtab{3}{
\begin{landscape}
\scriptsize
\begin{table}[!htb]
\caption{Atmospheric parameters for all stars analysed in the 51
  clusters: T$_{eff}$, log($g$), [Fe/H], [Mg/Fe] and
  [$\alpha$/Fe]. Membership identification is copied from Table
  \ref{starinfo} to guide the reader.}
\label{finalparam} 
\begin{tabular}{l|cc|l|cc|l|cc|c|cc|c}
\hline\hline
\noalign{\smallskip}
 {\rm NGC ID} & T$_{eff}^{(a)}$ (K) & T$_{eff}^{(b)}$ (K) & T$_{eff}^{(avg)}$ (K) & log($g$)$^{(a)}$ & log($g$)$^{(b)}$ & log($g$)$^{(avg)}$ & [Fe/H]$^{(a)}$ & [Fe/H]$^{(b)}$ & [Fe/H]$^{(avg)}$ & [Mg/Fe]$^{(a)}$ &  [$\alpha$/Fe]$^{(b)}$ & members\\
\noalign{\smallskip}
\hline
\noalign{\smallskip}
47Tuc\_502  &    3640$\pm$100  &  3627$\pm$125  &        3635$\pm$ 78  &   0.70$\pm$0.2    &  3.5$\pm$0.5         &     1.08$\pm$0.19    &  -0.10$\pm$0.07    &     -0.04$\pm$0.25    &    -0.10$\pm$0.07    &    0.23$\pm$0.1       &     0.37$\pm$0.05     &                  \\
47Tuc\_509  &    3983$\pm$84   &  3773$\pm$284  &        3966$\pm$ 81  &   1.47$\pm$0.30   &  1.1$\pm$0.9         &     1.43$\pm$0.28    &  -0.31$\pm$0.20    &     -1.60$\pm$0.44    &    -0.53$\pm$0.18    &    0.12$\pm$0.18      &     0.32$\pm$0.10     &       M       \\
47Tuc\_514  &    4651$\pm$217  &  4823$\pm$196  &        4746$\pm$145  &    2.3$\pm$0.5    &  2.9$\pm$0.5         &     2.61$\pm$0.36    &  -0.47$\pm$0.24    &     -0.75$\pm$0.25    &    -0.60$\pm$0.17    &    0.25$\pm$0.20      &     0.28$\pm$0.10     &       M       \\
47Tuc\_517  &    4688$\pm$224  &  4728$\pm$175  &        4713$\pm$138  &    2.4$\pm$0.5    &  2.6$\pm$0.4         &     2.51$\pm$0.33    &  -0.47$\pm$0.25    &     -0.90$\pm$0.20    &    -0.73$\pm$0.16    &    0.26$\pm$0.20      &     0.30$\pm$0.08     &       M       \\
47Tuc\_519  &    4263$\pm$154  &  4277$\pm$134  &        4271$\pm$101  &    1.8$\pm$0.4    &  1.96$\pm$0.35       &     1.89$\pm$0.26    &  -0.46$\pm$0.22    &     -1.05$\pm$0.15    &    -0.86$\pm$0.12    &    0.21$\pm$0.21      &     0.23$\pm$0.13     &       M       \\
47Tuc\_525  &    4272$\pm$152  &  4277$\pm$134  &        4275$\pm$101  &    1.8$\pm$0.4    &  1.95$\pm$0.35       &     1.88$\pm$0.26    &  -0.47$\pm$0.22    &     -1.10$\pm$0.20    &    -0.81$\pm$0.15    &    0.22$\pm$0.20      &     0.27$\pm$0.11     &       M       \\
47Tuc\_533  &    4278$\pm$107  &  4300$\pm$147  &        4286$\pm$ 86  &    1.8$\pm$0.4    &  1.8$\pm$0.4         &     1.82$\pm$0.28    &  -0.51$\pm$0.23    &     -0.94$\pm$0.27    &    -0.69$\pm$0.18    &    0.29$\pm$0.20      &     0.29$\pm$0.09     &       M       \\
47Tuc\_534  &    4879$\pm$363  &  5125$\pm$167  &        5082$\pm$152  &    2.3$\pm$0.8    &  2.8$\pm$0.4         &     2.70$\pm$0.36    &  -0.7 $\pm$0.4     &     -0.80$\pm$0.25           &    -0.77$\pm$0.21    &    0.29$\pm$0.15      &     0.19$\pm$0.12         &       M       \\
47Tuc\_535  &    4582$\pm$175  &  4596$\pm$121  &        4591$\pm$100  &    2.3$\pm$0.4    &  2.3$\pm$0.4         &     2.32$\pm$0.30    &  -0.46$\pm$0.19    &     -0.81$\pm$0.24    &    -0.59$\pm$0.15    &    0.25$\pm$0.21      &     0.23$\pm$0.11     &       M       \\
47Tuc\_539  &    4488$\pm$180  &  4628$\pm$125  &        4582$\pm$103  &    2.1$\pm$0.5    &  2.49$\pm$0.31       &     2.38$\pm$0.26    &  -0.46$\pm$0.21    &     -0.74$\pm$0.25    &    -0.58$\pm$0.16    &    0.25$\pm$0.21      &     0.25$\pm$0.08     &       M       \\
47Tuc\_551  &    4949$\pm$309  &  5148$\pm$123  &        5121$\pm$114  &    2.5$\pm$0.6    &  2.8$\pm$0.4         &     2.71$\pm$0.33    &  -0.61$\pm$0.29    &     -0.70$\pm$0.25    &    -0.66$\pm$0.19    &    0.30$\pm$0.13      &     0.20$\pm$0.14     &       M       \\
47Tuc\_553  &    4530$\pm$187  &  4623$\pm$125  &        4594$\pm$104  &    2.2$\pm$0.5    &  2.42$\pm$0.35       &     2.35$\pm$0.29    &  -0.47$\pm$0.18    &     -0.75$\pm$0.25    &    -0.57$\pm$0.15    &    0.24$\pm$0.20      &     0.22$\pm$0.10     &       M       \\
47Tuc\_554  &    3909$\pm$34   &  3750$\pm$156  &        3902$\pm$ 33  &   1.43$\pm$0.17   &  0.7$\pm$0.5         &     1.35$\pm$0.16    &  -0.28$\pm$0.19    &     -1.40$\pm$0.20    &    -0.81$\pm$0.14    &    0.12$\pm$0.20      &     0.26$\pm$0.11     &       M       \\
47Tuc\_559  &    4780$\pm$282  &  4749$\pm$193  &        4759$\pm$159  &    2.5$\pm$0.6    &  2.85$\pm$0.32       &     2.77$\pm$0.28    &  -0.52$\pm$0.32    &     -1.05$\pm$0.35    &    -0.76$\pm$0.24    &    0.27$\pm$0.20      &     0.32$\pm$0.10     &       M       \\
47Tuc\_571  &    4972$\pm$439  &  5277$\pm$260  &        5198$\pm$224  &    2.4$\pm$0.9    &  2.8$\pm$0.5         &     2.72$\pm$0.41    &  -0.7 $\pm$0.5     &     -0.80$\pm$0.33           &    -0.77$\pm$0.28    &    0.25$\pm$0.17      &     0.26$\pm$0.11         &       M       \\
47Tuc\_581  &    4868$\pm$346  &  5071$\pm$113  &        5051$\pm$107  &    2.2$\pm$0.8    &  2.44$\pm$0.34       &     2.40$\pm$0.31    &  -0.8 $\pm$0.4     &     -0.91$\pm$0.19           &    -0.89$\pm$0.17    &    0.36$\pm$0.13      &     0.22$\pm$0.15         &       M       \\
\noalign{\smallskip}
2298\_11   &   4766$\pm$339   &   4750$\pm$100  &    4751$\pm$ 96     &    1.70$\pm$0.60   &   1.90$\pm$0.37   &     1.84$\pm$0.31     &    -1.64$\pm$0.34    &    -2.00$\pm$0.10   &    -1.97$\pm$0.10   &    0.43$\pm$0.15   &    0.19$\pm$0.14  &     M    \\
2298\_14   &   4694$\pm$329   &   4775$\pm$ 74  &    4771$\pm$ 73     &    1.60$\pm$0.60   &   1.70$\pm$0.33   &     1.68$\pm$0.29     &    -1.63$\pm$0.32    &    -1.95$\pm$0.15   &    -1.89$\pm$0.14   &    0.45$\pm$0.13   &    0.20$\pm$0.14  &     M    \\
2298\_16   &   5006$\pm$336   &   5250$\pm$100  &    5230$\pm$ 96     &    2.20$\pm$0.80   &   2.30$\pm$0.33   &     2.29$\pm$0.31     &    -1.80$\pm$0.40    &    -2.00$\pm$0.10   &    -1.99$\pm$0.10   &    0.40$\pm$0.18   &    0.19$\pm$0.14  &     M    \\
2298\_17   &   4589$\pm$327   &   4799$\pm$188  &    4747$\pm$163     &    1.30$\pm$0.60   &   2.05$\pm$0.52   &     1.73$\pm$0.39     &    -1.68$\pm$0.35    &    -1.80$\pm$0.24   &    -1.76$\pm$0.20   &    0.47$\pm$0.12   &    0.23$\pm$0.13  &     M    \\
2298\_18   &   4894$\pm$344   &   5000$\pm$100  &    4992$\pm$ 96     &    2.00$\pm$0.70   &   1.95$\pm$0.41   &     1.96$\pm$0.35     &    -1.70$\pm$0.40    &    -2.00$\pm$0.10   &    -1.98$\pm$0.10   &    0.42$\pm$0.14   &    0.14$\pm$0.11  &     M    \\
\noalign{\smallskip}
 2808\_1   &   4676$\pm$331   &   4925$\pm$115  &   4898$\pm$108  &    1.80$\pm$0.90   &   1.65$\pm$0.23   &   1.66$\pm$0.22   &    -1.00$\pm$0.50    &    -1.15$\pm$0.23   &  -1.12$\pm$0.21  &    0.32$\pm$0.17   &    0.16$\pm$0.12  &     M    \\
 2808\_2   &   4736$\pm$167   &   5124$\pm$125  &   4985$\pm$100  &    2.52$\pm$0.30   &   3.25$\pm$0.25   &   2.95$\pm$0.19   &    -0.30$\pm$0.22    &    -0.25$\pm$0.25   &  -0.28$\pm$0.17  &    0.16$\pm$0.19   &    0.21$\pm$0.09  &         \\
 2808\_3   &   4812$\pm$360   &   4825$\pm$115  &   4824$\pm$109  &    2.20$\pm$0.90   &   2.50$\pm$0.39   &   2.45$\pm$0.36   &    -0.90$\pm$0.50    &    -1.35$\pm$0.23   &  -1.27$\pm$0.21  &    0.34$\pm$0.17   &    0.24$\pm$0.12  &     M    \\
 2808\_4   &   4949$\pm$314   &   4800$\pm$245  &   4856$\pm$193  &    2.50$\pm$0.70   &   2.00$\pm$0.45   &   2.15$\pm$0.38   &    -0.57$\pm$0.30    &    -1.20$\pm$0.24   &  -0.95$\pm$0.19  &    0.24$\pm$0.13   &    0.24$\pm$0.14  &     M    \\
 2808\_5   &   4801$\pm$399   &   5276$\pm$208  &   5175$\pm$184  &    1.90$\pm$0.80   &   2.65$\pm$0.45   &   2.47$\pm$0.39   &    -1.40$\pm$0.40    &    -1.20$\pm$0.24   &  -1.25$\pm$0.21  &    0.42$\pm$0.16   &    0.19$\pm$0.14  &     M    \\
 2808\_6   &   4910$\pm$354   &   4750$\pm$158  &   4777$\pm$144  &    2.50$\pm$0.90   &   2.20$\pm$0.56   &   2.28$\pm$0.48   &    -0.70$\pm$0.50    &    -1.40$\pm$0.20   &  -1.30$\pm$0.19  &    0.32$\pm$0.20   &    0.26$\pm$0.09  &     M    \\
 2808\_7   &   4903$\pm$401   &   4799$\pm$149  &   4812$\pm$140  &    2.30$\pm$0.90   &   1.85$\pm$0.32   &   1.90$\pm$0.30   &    -0.80$\pm$0.50    &    -1.45$\pm$0.15   &  -1.40$\pm$0.14  &    0.31$\pm$0.20   &    0.19$\pm$0.14  &     M    \\
 2808\_8   &   4786$\pm$413   &   4651$\pm$122  &   4662$\pm$117  &    2.30$\pm$0.80   &   1.40$\pm$0.49   &   1.65$\pm$0.42   &    -0.51$\pm$0.34    &    -1.20$\pm$0.24   &  -0.97$\pm$0.20  &    0.12$\pm$0.06   &    0.21$\pm$0.14  &     M    \\
 2808\_9   &   4687$\pm$362   &   4925$\pm$114  &   4904$\pm$109  &    1.70$\pm$0.80   &   2.00$\pm$0.45   &   1.93$\pm$0.39   &    -1.40$\pm$0.40    &    -1.50$\pm$0.10   &  -1.49$\pm$0.10  &    0.42$\pm$0.14   &    0.22$\pm$0.12  &        \\
2808\_10   &   4072$\pm$114   &   3926$\pm$296  &   4053$\pm$106  &    1.40$\pm$0.40   &   0.90$\pm$0.73   &   1.28$\pm$0.35   &    -0.47$\pm$0.22    &    -1.50$\pm$0.39   &  -0.72$\pm$0.19  &    0.19$\pm$0.24   &    0.32$\pm$0.07  &     M    \\
2808\_11   &   4662$\pm$415   &   4900$\pm$122  &   4881$\pm$117  &    1.90$\pm$0.90   &   2.05$\pm$0.35   &   2.03$\pm$0.33   &    -0.80$\pm$0.50    &    -1.00$\pm$0.10   &  -0.99$\pm$0.10  &    0.30$\pm$0.17   &    0.20$\pm$0.12  &        \\
2808\_13   &   4961$\pm$279   &   4773$\pm$236  &   4851$\pm$180  &    2.70$\pm$0.60   &   1.90$\pm$0.37   &   2.12$\pm$0.31   &    -0.52$\pm$0.31    &    -1.25$\pm$0.25   &  -0.96$\pm$0.19  &    0.24$\pm$0.17   &    0.25$\pm$0.13  &     M    \\
2808\_14   &   4968$\pm$300   &   5049$\pm$ 99  &   5041$\pm$ 94  &    2.50$\pm$0.60   &   2.40$\pm$0.37   &   2.43$\pm$0.31   &    -0.57$\pm$0.31    &    -0.90$\pm$0.20   &  -0.80$\pm$0.17  &    0.25$\pm$0.13   &    0.19$\pm$0.14  &     M    \\
2808\_15   &   4638$\pm$430   &   5075$\pm$225  &   4981$\pm$199  &    1.50$\pm$0.80   &   2.00$\pm$0.39   &   1.90$\pm$0.35   &    -1.20$\pm$0.50    &    -1.15$\pm$0.23   &  -1.16$\pm$0.21  &    0.38$\pm$0.18   &    0.22$\pm$0.14  &     M    \\
2808\_16   &   4328$\pm$ 48   &   4526$\pm$175  &   4342$\pm$ 46  &    1.20$\pm$0.05   &   1.90$\pm$0.44   &   1.21$\pm$0.05   &    -0.72$\pm$0.24    &    -1.10$\pm$0.20   &  -0.94$\pm$0.15  &    0.15$\pm$0.14   &    0.22$\pm$0.11  &     M    \\
2808\_17   &   4832$\pm$264   &   4625$\pm$125  &   4663$\pm$113  &    2.50$\pm$0.70   &   2.05$\pm$0.27   &   2.11$\pm$0.25   &    -0.51$\pm$0.29    &    -1.25$\pm$0.25   &  -0.93$\pm$0.19  &    0.19$\pm$0.15   &    0.25$\pm$0.10  &     M    \\
2808\_18   &   4433$\pm$ 74   &   4576$\pm$159  &   4458$\pm$ 67  &    2.36$\pm$0.23   &   2.55$\pm$0.47   &   2.40$\pm$0.21   &    -0.45$\pm$0.08    &    -0.85$\pm$0.32   &  -0.47$\pm$0.08  &    0.16$\pm$0.15   &    0.27$\pm$0.11  &         \\
2808\_19   &   4874$\pm$359   &   5175$\pm$114  &   5147$\pm$109  &    2.30$\pm$0.90   &   3.15$\pm$0.32   &   3.05$\pm$0.30   &    -1.00$\pm$0.50    &    -1.05$\pm$0.15   &  -1.05$\pm$0.14  &    0.34$\pm$0.19   &    0.18$\pm$0.14  &     M    \\
\noalign{\smallskip}
...  &   ...  &  ...    &... &    ...   & ...    &    ...    &  ...   &   ...       &    ...   &    ...      &     ...    &     ...     \\
\noalign{\smallskip}
\hline
\end{tabular}
\tablefoot{Complete version of this Table for all 758 stars is
  available online at VizieR.}
\tablefoottext{a}{Results using only MILES library.}
\tablefoottext{b}{Results using only COELHO library.}
\tablefoottext{avg}{Average of (a) and (b) results.}
\end{table}
\end{landscape}
}
\normalsize

\longtab{4}{
\footnotesize
\begin{longtable}{llrcccccccc}
\caption{\label{finalparamavgcalib} Final parameters for the 51 clusters ${\rm [Fe/H]}$, ${\rm
      [Mg/Fe]}$, ${\rm [\alpha/Fe]}$ and $v_{\rm helio}$. Columns
    labelled with `MILES', `Coelho', and $v_{\rm helio}$ are averages
    of individual stars from Tables \ref{starinfo} and
    \ref{finalparam}. `Car09' are the metallicities from the
    compilation of \cite{carretta+09}, identified accordingly with
      their Table A.1 with (1), if it is the average of different
      metallicity scales or with (2) if it is the value from Harris
      catalogue plus an offset. The adopted value of [Fe/H] is
    marked with an asterisk. For this column the error comes from the formal error 
    propagation of the average, and the value in brackets is the standard deviation 
    of MILES and Coelho values.
      The systematic differences between [Fe/H]$_{\rm avg}$ and [Fe/H]$_{\rm C09}$
      are null, as shown in Fig. \ref{d14xc09-51}.}\\
\hline\hline
\noalign{\smallskip}
Cluster  & Other  & $v_{\rm helio}$ &  [Fe/H]   &  [Fe/H]      & [Fe/H]*     & [Fe/H]   & [Mg/Fe] & [$\alpha$/Fe]\\
             & names &  (km/s)         &  (MILES)  & (Coelho)  & (average) & (Car09)  & (MILES)  & (Coelho) \\
\noalign{\smallskip}
\hline
\noalign{\smallskip}
\endfirsthead
\caption[]{continued.}\\
\hline\hline
\noalign{\smallskip}
Cluster  & Other  & $v_{\rm helio}$ &  [Fe/H]   &  [Fe/H]      & [Fe/H]*     & [Fe/H]   & [Mg/Fe] & [$\alpha$/Fe]\\
             & names &  (km/s)         &  (MILES)  & (Coelho)  & (average) & (Car09)  & (MILES)  & (Coelho) \\
\noalign{\smallskip}
\hline
\noalign{\smallskip}
\endhead
\hline
\endfoot
NGC 104  &      47 Tuc           &     -48$\pm$10     &     -0.46$\pm$0.06    &   -0.95$\pm$0.06    &   -0.71$\pm$0.04 [0.35]    &    -0.76$\pm$0.02 (1)  &  0.26$\pm$0.05     &   0.26$\pm$0.03     \\ 
NGC 2298 &                       &     134$\pm$14     &     -1.68$\pm$0.16    &   -1.98$\pm$0.05    &   -1.95$\pm$0.05 [0.21]    &    -1.96$\pm$0.04 (1)  &  0.44$\pm$0.06     &   0.19$\pm$0.06     \\ 
NGC 2808 &                       &      96$\pm$25     &     -0.67$\pm$0.09    &   -1.21$\pm$0.06    &   -1.06$\pm$0.05 [0.38]    &    -1.18$\pm$0.04 (1)  &  0.22$\pm$0.04     &   0.24$\pm$0.03     \\ 
NGC 3201 &                       &     472$\pm$19     &     -1.52$\pm$0.08    &   -1.51$\pm$0.04    &   -1.51$\pm$0.03 [0.01]    &    -1.51$\pm$0.02 (1)  &  0.43$\pm$0.04         &   0.22$\pm$0.03         \\ 
NGC 4372 &                       &      63$\pm$10     &     -1.83$\pm$0.12    &   -2.34$\pm$0.07    &   -2.2[2$\pm$0.06 [0.36]    &    -2.19$\pm$0.08 (1)  &  0.40$\pm$0.04     &   0.21$\pm$0.04     \\ 
Rup 106  &                       &     -47$\pm$12     &     -1.60$\pm$0.12    &   -1.54$\pm$0.05    &   -1.54$\pm$0.04 ][0.04]    &    -1.78$\pm$0.08 (1)  &  0.45$\pm$0.05     &   0.12$\pm$0.03     \\ 
NGC 4590 &      M 68         &     -92$\pm$25     &     -1.86$\pm$0.15    &   -2.23$\pm$0.05    &   -2.20$\pm$0.05  [0.26]     &    -2.27$\pm$0.04 (1)  &  0.39$\pm$0.05     &   0.19$\pm$0.05     \\ 
NGC 5634 &               &         -29$\pm$42     &     -1.60$\pm$0.10    &   -1.82$\pm$0.07    &   -1.75$\pm$0.06 [0.16]     &    -1.93$\pm$0.09 (2)  &  0.43$\pm$0.05     &   0.20$\pm$0.04     \\ 
NGC 5694 &               &        -150$\pm$ 9     &     -1.76$\pm$0.12    &   -2.00$\pm$0.04    &   -1.98$\pm$0.04 [0.17]    &    -2.02$\pm$0.07 (1)  &  0.41$\pm$0.05         &   0.17$\pm$0.04     \\ 
NGC 5824 &                       &     -35$\pm$12     &     -1.95$\pm$0.05    &   -2.01$\pm$0.03    &   -1.99$\pm$0.03 [0.04]    &    -1.94$\pm$0.14 (1)  &  0.44$\pm$0.03     &   0.24$\pm$0.03     \\ 
NGC 5897 &                       &      88$\pm$13     &     -1.63$\pm$0.12    &   -2.02$\pm$0.04    &   -1.97$\pm$0.04 [0.28]    &    -1.90$\pm$0.06 (1)  &  0.43$\pm$0.05     &   0.23$\pm$0.04     \\ 
NGC 5904 &      M 5                  &      46$\pm$ 7     &     -1.02$\pm$0.13    &   -1.30$\pm$0.06    &   -1.25$\pm$0.05 [0.20]    &    -1.33$\pm$0.02 (1)  &  0.35$\pm$0.05     &   0.24$\pm$0.04     \\ 
NGC 5927 &               &     -99$\pm$11     &     -0.16$\pm$0.03    &   -0.82$\pm$0.10    &   -0.21$\pm$0.02 [0.46]    &    -0.29$\pm$0.07 (1)  &  0.09$\pm$0.04          &   0.30$\pm$0.04     \\ 
NGC 5946 &                       &     134$\pm$29     &     -1.50$\pm$0.17    &   -1.55$\pm$0.07    &   -1.54$\pm$0.07 [0.04]    &    -1.29$\pm$0.14 (1)  &  0.42$\pm$0.07     &   0.22$\pm$0.05     \\ 
BH 176   &                       &      -6$\pm$14     &     -0.08$\pm$0.04    &   -0.07$\pm$0.06    &   -0.08$\pm$0.04 [0.01]    &         ---            &  0.10$\pm$0.05    &   0.20$\pm$0.03     \\ 
Lynga 7  &      BH 184           &     -13$\pm$28     &     -0.41$\pm$0.13    &   -0.87$\pm$0.15    &   -0.61$\pm$0.10 [0.33]    &         ---            &  0.21$\pm$0.11    &   0.27$\pm$0.06     \\ 
Pal 14   &      AvdB             &      45$\pm$ 9     &     -1.07$\pm$0.18    &   -1.27$\pm$0.10    &   -1.22$\pm$0.09 [0.14]    &    -1.63$\pm$0.08 (1)  &  0.32$\pm$0.06     &   0.24$\pm$0.04     \\ 
NGC 6121 &      M 4                  &      23$\pm$35     &     -0.80$\pm$0.13    &   -1.05$\pm$0.06    &   -1.01$\pm$0.05 [0.18]    &    -1.18$\pm$0.02 (1)  &  0.35$\pm$0.06     &   0.27$\pm$0.04     \\ 
NGC 6171 &      M 107        &    -122$\pm$ 0     &     -0.55$\pm$0.27    &   -1.00$\pm$0.10    &   -0.95$\pm$0.09 [0.32]    &    -1.03$\pm$0.02 (1)  &  0.28$\pm$0.21         &   0.20$\pm$0.14     \\ 
NGC 6254 &      M 10         &      43$\pm$34     &     -1.59$\pm$0.08    &   -1.55$\pm$0.04    &   -1.56$\pm$0.04 [0.03]    &    -1.57$\pm$0.02 (1)  &  0.44$\pm$0.03         &   0.21$\pm$0.03     \\ 
NGC 6284 &                       &      35$\pm$21     &     -0.84$\pm$0.15    &   -1.12$\pm$0.07    &   -1.07$\pm$0.06 [0.20]    &    -1.31$\pm$0.09 (2)  &  0.33$\pm$0.06     &   0.27$\pm$0.04     \\ 
NGC 6316 &                       &      81$\pm$40     &     -0.28$\pm$0.08    &   -0.84$\pm$0.10    &   -0.50$\pm$0.06 [0.40]    &    -0.36$\pm$0.14 (1)  &  0.11$\pm$0.06     &   0.30$\pm$0.03     \\ 
NGC 6356 &                       &      36$\pm$41     &     -0.30$\pm$0.06    &   -0.91$\pm$0.07    &   -0.55$\pm$0.04 [0.43]    &    -0.35$\pm$0.14 (1)  &  0.12$\pm$0.04     &   0.30$\pm$0.02     \\ 
NGC 6355 &                       &    -215$\pm$ 7     &     -1.38$\pm$0.09    &   -1.54$\pm$0.08    &   -1.46$\pm$0.06 [0.11]    &    -1.33$\pm$0.14 (1)  &  0.26$\pm$0.01     &   0.27$\pm$0.05     \\ 
NGC 6352 &                       &    -138$\pm$36     &     -0.41$\pm$0.06    &   -0.76$\pm$0.07    &   -0.54$\pm$0.04 [0.25]    &    -0.62$\pm$0.05 (1)  &  0.15$\pm$0.05     &   0.30$\pm$0.02     \\ 
NGC 6366 &                       &    -137$\pm$54     &     -0.41$\pm$0.07    &   -0.81$\pm$0.07    &   -0.61$\pm$0.05 [0.28]    &    -0.59$\pm$0.08 (1)  &  0.22$\pm$0.05     &   0.30$\pm$0.02     \\ 
HP 1     &      BH 229       &      54$\pm$ 5     &     -0.70$\pm$0.11    &   -1.49$\pm$0.09    &   -1.17$\pm$0.07 [0.56]    &    -1.57$\pm$0.09 (2)  &  0.33$\pm$0.07         &   0.28$\pm$0.04     \\ 
NGC 6401 &                       &    -120$\pm$17     &     -0.59$\pm$0.14    &   -1.34$\pm$0.09    &   -1.12$\pm$0.07 [0.53]    &    -1.01$\pm$0.14 (1)  &  0.32$\pm$0.08     &   0.27$\pm$0.04     \\ 
NGC 6397 &                       &     -27$\pm$55     &     -1.75$\pm$0.06    &   -2.15$\pm$0.03    &   -2.07$\pm$0.03 [0.29]    &    -1.99$\pm$0.02 (1)  &  0.40$\pm$0.03     &   0.23$\pm$0.03     \\ 
Pal 6    &                   &     177$\pm$ 5     &     -0.27$\pm$0.14    &   -1.66$\pm$0.17    &   -0.85$\pm$0.11 [0.98]    &    -1.06$\pm$0.09 (2)  &  0.14$\pm$0.10         &   0.28$\pm$0.05     \\ 
NGC 6426 &                       &    -242$\pm$11     &     -2.03$\pm$0.11    &   -2.46$\pm$0.05    &   -2.39$\pm$0.04 [0.30]    &         ---            &  0.38$\pm$0.06    &   0.24$\pm$0.05     \\ 
NGC 6440 &                       &     -59$\pm$26     &     -0.03$\pm$0.06    &   -0.80$\pm$0.10    &   -0.24$\pm$0.05 [0.54]    &    -0.20$\pm$0.14 (1)  &  0.11$\pm$0.04     &   0.31$\pm$0.03     \\ 
NGC 6441 &                       &      -6$\pm$32     &     -0.18$\pm$0.09    &   -0.71$\pm$0.10    &   -0.41$\pm$0.07 [0.37]    &    -0.44$\pm$0.07 (1)  &  0.11$\pm$0.06     &   0.26$\pm$0.04     \\ 
NGC 6453 &                       &    -153$\pm$11     &     -1.45$\pm$0.18    &   -1.57$\pm$0.10    &   -1.54$\pm$0.09 [0.08]    &    -1.48$\pm$0.14 (1)  &  0.42$\pm$0.09     &   0.16$\pm$0.06     \\ 
Djorg 2  &      ESO456SC38   &    -150$\pm$28     &     -0.50$\pm$0.12    &   -1.19$\pm$0.14    &   -0.79$\pm$0.09 [0.49]    &         ---            &  0.28$\pm$0.10      &   0.27$\pm$0.05     \\ 
NGC 6528 &                       &     185$\pm$10     &     -0.07$\pm$0.10    &   -0.18$\pm$0.08    &   -0.13$\pm$0.07 [0.08]    &    +0.07$\pm$0.08 (1)  &  0.05$\pm$0.09     &   0.26$\pm$0.05     \\ 
NGC 6539 &                       &      30$\pm$18     &     -0.23$\pm$0.09    &   -0.89$\pm$0.09    &   -0.55$\pm$0.06 [0.47]    &    -0.53$\pm$0.14 (1)  &  0.16$\pm$0.07     &   0.30$\pm$0.03     \\ 
NGC 6553 &                       &       6$\pm$ 8     &     -0.12$\pm$0.01    &   -0.55$\pm$0.07    &   -0.13$\pm$0.01 [0.30]    &    -0.16$\pm$0.06 (1)  &  0.11$\pm$0.01     &   0.30$\pm$0.02     \\ 
NGC 6558 &                       &    -210$\pm$16     &     -0.88$\pm$0.20    &   -1.02$\pm$0.05    &   -1.01$\pm$0.05 [0.10]    &    -1.37$\pm$0.14 (1)  &  0.26$\pm$0.06     &   0.23$\pm$0.06     \\ 
IC 1276  &      Pal 7        &     155$\pm$15     &     -0.13$\pm$0.06    &   -1.11$\pm$0.07    &   -0.56$\pm$0.05 [0.69]    &    -0.65$\pm$0.09 (2)  &  0.09$\pm$0.05         &   0.30$\pm$0.03     \\ 
NGC 6569 &                       &     -51$\pm$ 9     &     -0.53$\pm$0.09    &   -0.85$\pm$0.11    &   -0.66$\pm$0.07 [0.23]    &    -0.72$\pm$0.14 (1)  &  0.30$\pm$0.07     &   0.29$\pm$0.03     \\ 
NGC 6656 &      M 22         &    -152$\pm$25     &     -1.77$\pm$0.05    &   -1.94$\pm$0.02    &   -1.92$\pm$0.02 [0.12]    &    -1.70$\pm$0.08 (1)  &  0.50$\pm$0.01         &   0.22$\pm$0.02     \\ 
NGC 6749 &                       &     -66$\pm$ 8     &     -0.64$\pm$0.15    &   -2.14$\pm$0.11    &   -1.59$\pm$0.09 [1.06]    &    -1.62$\pm$0.09 (2)  &  0.34$\pm$0.10     &   0.17$\pm$0.06     \\ 
NGC 6752 &                       &     -28$\pm$ 7     &     -1.49$\pm$0.13    &   -1.59$\pm$0.08    &   -1.57$\pm$0.07 [0.07]    &    -1.55$\pm$0.01 (1)  &  0.47$\pm$0.06     &   0.22$\pm$0.05     \\ 
Pal 10   &                       &     -38$\pm$17     &     -0.08$\pm$0.04    &   -0.53$\pm$0.05    &   -0.24$\pm$0.03 [0.32]    &         ---            &  0.12$\pm$0.01    &   0.27$\pm$0.03     \\ 
Terzan 8 &                       &     135$\pm$19     &     -1.76$\pm$0.07    &   -2.18$\pm$0.05    &   -2.06$\pm$0.04 [0.30]    &         ---            &  0.41$\pm$0.04    &   0.21$\pm$0.04     \\ 
Pal 11   &                       &     -81$\pm$15     &     -0.22$\pm$0.05    &   -0.62$\pm$0.08    &   -0.35$\pm$0.05 [0.28]    &    -0.45$\pm$0.08 (1)  &  0.12$\pm$0.05     &   0.30$\pm$0.03     \\ 
NGC 6838 &      M 71             &     -42$\pm$18     &     -0.48$\pm$0.08    &   -0.77$\pm$0.08    &   -0.63$\pm$0.06 [0.21]    &    -0.82$\pm$0.02 (1)  &  0.25$\pm$0.07     &   0.29$\pm$0.03     \\ 
NGC 6864 &      M 75             &    -190$\pm$20     &     -0.75$\pm$0.10    &   -1.09$\pm$0.06    &   -1.00$\pm$0.05 [0.24]    &    -1.29$\pm$0.14 (1)  &  0.35$\pm$0.05     &   0.22$\pm$0.03     \\ 
NGC 7006 &                       &    -391$\pm$24     &     -1.54$\pm$0.19    &   -1.74$\pm$0.11    &   -1.69$\pm$0.09 [0.14]    &    -1.46$\pm$0.06 (1)  &  0.42$\pm$0.07     &   0.25$\pm$0.05     \\ 
NGC 7078 &      M 15         &    -159$\pm$40     &     -2.11$\pm$0.02    &   -2.49$\pm$0.03    &   -2.23$\pm$0.02 [0.26]    &    -2.33$\pm$0.02 (1)  &  0.41$\pm$0.03         &   0.24$\pm$0.03     \\ 
\noalign{\smallskip}
\end{longtable}
}
\normalsize

\end{document}